\input harvmac
\noblackbox

\let\includefigures=\iftrue
\let\useblackboard=\iftrue
\newfam\black

%Figure Stuff
\includefigures
\message{If you do not have epsf.tex (to include figures),}
\message{change the option at the top of the tex file.}
\input epsf
\def\figin{\epsfcheck\figin}\def\figins{\epsfcheck\figins}
\def\epsfcheck{\ifx\epsfbox\UnDeFiNeD
\message{(NO epsf.tex, FIGURES WILL BE IGNORED)}
\gdef\figin##1{\vskip2in}\gdef\figins##1{\hskip.5in}% blank space instead
\else\message{(FIGURES WILL BE INCLUDED)}%
\gdef\figin##1{##1}\gdef\figins##1{##1}\fi}
\def\DefWarn#1{}
\def\figinsert{\goodbreak\midinsert}
\def\ifig#1#2#3{\DefWarn#1\xdef#1{Fig.~\the\figno}
\writedef{#1\leftbracket Fig.\noexpand~\the\figno}%
\figinsert\figin{\centerline{#3}}\medskip\centerline{\vbox{
\baselineskip12pt\advance\hsize by -1truein
\noindent\footnotefont{\bf Fig.~\the\figno:} #2}}
\bigskip\endinsert\global\advance\figno by1}
%%%
\else
\def\ifig#1#2#3{\xdef#1{Fig.~\the\figno}
\writedef{#1\leftbracket Fig.\noexpand~\the\figno}%
%\figinsert\figin{\centerline{#3}}\medskip
%\centerline{\vbox{\baselineskip12pt
%\advance\hsize by -1truein\noindent
%\footnotefont{\bf Fig.~\the\figno:} #2}}
%\bigskip\endinsert
\global\advance\figno by1}
\fi

\def\doublefig#1#2#3#4{\DefWarn#1\xdef#1{Fig.~\the\figno}
\writedef{#1\leftbracket Fig.\noexpand~\the\figno}%
\figinsert\figin{\centerline{#3\hskip1.0cm#4}}\medskip\centerline{\vbox{
\baselineskip12pt\advance\hsize by -1truein
\noindent\footnotefont{\bf Fig.~\the\figno:} #2}}
\bigskip\endinsert\global\advance\figno by1}

%%BLACKBOARD FONT STUFF
\useblackboard
\message{If you do not have msbm (blackboard bold) fonts,}
\message{change the option at the top of the tex file.}
\font\blackboard=msbm10 scaled \magstep1
\font\blackboards=msbm7
\font\blackboardss=msbm5
\textfont\black=\blackboard
\scriptfont\black=\blackboards
\scriptscriptfont\black=\blackboardss

\else

\fi
% *************************************
%\draft
%
\def\subsubsec#1{\bigskip\noindent{\it{#1}} \bigskip}
\def\yboxit#1#2{\vbox{\hrule height #1 \hbox{\vrule width #1
\vbox{#2}\vrule width #1 }\hrule height #1 }}
\def\fillbox#1{\hbox to #1{\vbox to #1{\vfil}\hfil}}
\def\ybox{{\lower 1.3pt \yboxit{0.4pt}{\fillbox{8pt}}\hskip-0.2pt}}
%
%
%%MATH MACROS
%Greek letters and their bars

%More bars

%\def\l{\left}
%\def\r{\right}
\def\comments#1{}

%AEL

%AEL

\def\II{\relax{I\kern-.10em I}}

\def\IZ{\relax{\rm Z\kern-.34em Z}}
\def\IB{\relax{\rm I\kern-.18em B}}
\def\IC{{\relax\hbox{$\inbar\kern-.3em{\rm C}$}}}
\def\ID{\relax{\rm I\kern-.18em D}}
\def\IE{\relax{\rm I\kern-.18em E}}
\def\IF{\relax{\rm I\kern-.18em F}}
\def\IG{\relax\hbox{$\inbar\kern-.3em{\rm G}$}}
\def\IGa{\relax\hbox{${\rm I}\kern-.18em\Gamma$}}
\def\IH{\relax{\rm I\kern-.18em H}}
\def\II{\relax{\rm I\kern-.18em I}}
\def\IK{\relax{\rm I\kern-.18em K}}
\def\IP{\relax{\rm I\kern-.18em P}}
%\def\IX{\relax{\rm X\kern-.01em X}}
%this doesn't work

%

\def\inbar{\,\vrule height1.5ex width.4pt depth0pt}

\def\IR{\relax{\rm I\kern-.18em R}}

\def\simgt{\hskip0.05in\relax{
\raise3.0pt\hbox{ $>$
{\lower5.0pt\hbox{\kern-1.05em $\sim$}} }} \hskip0.05in}

%

 % for now

%

\def\lp10{\ell_p^{10}}
\def\lp11{\ell_p^{11}}
\def\R11{R_{11}}

\def\frac#1#2{{#1 \over #2}}

%identity operator from doyon-fonseca

%% from the topological vertex paper

%%                              TABLEAUX.TEX
%%      This  macro file is for producing a ``Young Tableau'' which is
%%      an array of little squares sometimes used in mathematical physics.
%%      For instance, the command $\tableau{6 3 2}$ will produce a tableau
%%      with 6 squares in the top row, 3 in the next, and 2 in the last.
%%                                  OOOOOO
%%      This tableau will look like OOO    but made of squares instead of O's.
%%                                  OO
%%      Any number of rows may be present, each having a nonzero number of
%%      squares.
%%
%%      A tableau is math mode material, so use $ or $$ to enclose it.
%%
%%      The size and line-thickness of the little boxes are controlled by the
%%      dimension parameters --
%%              \tableauside=1.0ex              %(size)
%%              \tableaurule=0.4pt              %(line-thickness)
%%      Change them if you want.
%%
%%                                                      -- Doug Eardley 9/19/8%%
%%
\newdimen\tableauside\tableauside=1.0ex
\newdimen\tableaurule\tableaurule=0.4pt
\newdimen\tableaustep
\def\phantomhrule#1{\hbox{\vbox to0pt{\hrule height\tableaurule width#1\vss}}}
\def\phantomvrule#1{\vbox{\hbox to0pt{\vrule width\tableaurule height#1\hss}}}
\def\sqr{\vbox{%
  \phantomhrule\tableaustep
  \hbox{\phantomvrule\tableaustep\kern\tableaustep\phantomvrule\tableaustep}%
  \hbox{\vbox{\phantomhrule\tableauside}\kern-\tableaurule}}}
\def\squares#1{\hbox{\count0=#1\noindent\loop\sqr
  \advance\count0 by-1 \ifnum\count0>0\repeat}}
\def\tableau#1{\vcenter{\offinterlineskip
  \tableaustep=\tableauside\advance\tableaustep by-\tableaurule
  \kern\normallineskip\hbox
    {\kern\normallineskip\vbox
      {\gettableau#1 0 }%
     \kern\normallineskip\kern\tableaurule}%
  \kern\normallineskip\kern\tableaurule}}
\def\gettableau#1 {\ifnum#1=0\let\next=\null\else
  \squares{#1}\let\next=\gettableau\fi\next}

\tableauside=1.0ex
\tableaurule=0.4pt

%% from shiraz

 %
 %       \eqn\label{a+b=c}       gives displayed equation, numbered
 %                               consecutively within sections.
%     \eqnn and \eqna define labels in advance (of eqalign?)
 %
 \def\eqnn#1{\xdef #1{(\secsym\the\meqno)}\writedef{#1\leftbracket#1}%
 \global\advance\meqno by1\wrlabeL#1}
 \def\eqna#1{\xdef #1##1{\hbox{$(\secsym\the\meqno##1)$}}
 \writedef{#1\numbersign1\leftbracket#1{\numbersign1}}%
 \global\advance\meqno by1\wrlabeL{#1$\{\}$}}
 \def\eqn#1#2{\xdef #1{(\secsym\the\meqno)}\writedef{#1\leftbracket#1}%
 \global\advance\meqno by1$$#2\eqno#1\eqlabeL#1$$}

\global\newcount\itemno \global\itemno=0

%\bigbreak}
%\bigskip\noindent}
\def\itemaut#1{\global\advance\itemno by1\noindent\item{\the\itemno.}#1}

%\itemized
%\itemaut{First this.}
%\itemaut{Then that.}

%%ENGLISH MACROS
\def\eg{{\it e.g.}}

\hyphenation{Di-men-sion-al}

%%REFERENCING MACROS

%%

\lref\otherT{
%\CostaEJ
  M.~S.~Costa, C.~A.~R.~Herdeiro, J.~Penedones and N.~Sousa,
  ``Hagedorn transition and chronology protection in string theory,''
  Nucl.\ Phys.\ B {\bf 728}, 148 (2005)
  [arXiv:hep-th/0504102];
  %%CITATION = HEP-TH 0504102;%%
%\BerkoozYM
  M.~Berkooz, Z.~Komargodski, D.~Reichmann and V.~Shpitalnik,
  ``Flow of geometries and instantons on the null orbifold,''
  arXiv:hep-th/0507067;
  %%CITATION = HEP-TH 0507067;%%
%\YangRX
  H.~Yang and B.~Zwiebach,
  ``A closed string tachyon vacuum?,''
  JHEP {\bf 0509}, 054 (2005)
  [arXiv:hep-th/0506077];
  %%CITATION = HEP-TH 0506077;%%
%\KruczenskiPJ
  M.~Kruczenski and A.~Lawrence,
  ``Random walks and the Hagedorn transition,''
  arXiv:hep-th/0508148;
  %%CITATION = HEP-TH 0508148;%%
 %\SilversteinQF
  E.~Silverstein,
  ``Dimensional mutation and spacelike singularities,''
  arXiv:hep-th/0510044;
  %%CITATION = HEP-TH 0510044;%%
  %\SheMT
  %\BergmanQF
  O.~Bergman and S.~Hirano,
  ``Semi-localized instability of the Kaluza-Klein linear dilaton vacuum,''
  arXiv:hep-th/0510076;
  %%CITATION = HEP-TH 0510076;%%
%\FreedmanWX
  D.~Z.~Freedman, M.~Headrick and A.~Lawrence,
  ``On closed string tachyon dynamics,''
  arXiv:hep-th/0510126;
  %%CITATION = HEP-TH 0510126;%%
  J.~L.~F.~Barbon and E.~Rabinovici,
  ``Remarks on black hole instabilities and closed string tachyons,''
  Found.\ Phys.\  {\bf 33}, 145 (2003)
  [arXiv:hep-th/0211212].
  %%CITATION = HEP-TH 0211212;%%
}

\lref\cosmoT{
H.~Yang and B.~Zwiebach,
  ``Rolling closed string tachyons and the big crunch,''
  JHEP {\bf 0508}, 046 (2005)
  [arXiv:hep-th/0506076];
  %%CITATION = HEP-TH 0506076;%%
J.~H.~She,
  ``A matrix model for Misner universe,''
  arXiv:hep-th/0509067;
  %%CITATION = HEP-TH 0509067;%%
%\HikidaXA
  Y.~Hikida and T.~S.~Tai,
  ``D-instantons and closed string tachyons in Misner space,''
  arXiv:hep-th/0510129;
  %%CITATION = HEP-TH 0510129;%%
}

%\StromingerPC
\lref\StromingerPC{
  A.~Strominger,
  ``Open string creation by S-branes,''
  arXiv:hep-th/0209090.
  %%CITATION = HEP-TH 0209090;%%
}

\lref\AdSCFTsing{ T.~Hertog and G.~T.~Horowitz,
``Holographic description of AdS cosmologies,''
  JHEP {\bf 0504}, 005 (2005)
  [arXiv:hep-th/0503071],
  %%CITATION = HEP-TH 0503071;%%
   ``Towards a big crunch dual,''
  JHEP {\bf 0407}, 073 (2004)
  [arXiv:hep-th/0406134];
  %%CITATION = HEP-TH 0406134;%%
L.~Fidkowski, V.~Hubeny, M.~Kleban and S.~Shenker,
  ``The black hole singularity in AdS/CFT,''
  JHEP {\bf 0402}, 014 (2004)
  [arXiv:hep-th/0306170];
  %%CITATION = HEP-TH 0306170;%%
  P.~Kraus, H.~Ooguri and S.~Shenker,
  ``Inside the horizon with AdS/CFT,''
  Phys.\ Rev.\ D {\bf 67}, 124022 (2003)
  [arXiv:hep-th/0212277];
  %%CITATION = HEP-TH 0212277;%%
  %\FestucciaPI
  G.~Festuccia and H.~Liu,
  ``Excursions beyond the horizon: Black hole singularities in Yang-Mills
  theories. I,''
  arXiv:hep-th/0506202;
  %%CITATION = HEP-TH 0506202;%%
  %\FuruuchiZP
  K.~Furuuchi,
  ``Confined phase in real time formalism and the fate of the world behind the
  horizon,''
  arXiv:hep-th/0510056.
  %%CITATION = HEP-TH 0510056;%%
  }

\lref\corrprinc{
%\SusskindWS
  L.~Susskind,
  ``Some speculations about black hole entropy in string theory,''
  arXiv:hep-th/9309145.
  %%CITATION = HEP-TH 9309145;%%
%\HorowitzNW
  G.~T.~Horowitz and J.~Polchinski,
  ``A correspondence principle for black holes and strings,''
  Phys.\ Rev.\ D {\bf 55}, 6189 (1997)
  [arXiv:hep-th/9612146].
  %%CITATION = HEP-TH 9612146;%%
}

%\PolyakovTP
\lref\PolyakovTP{
  A.~M.~Polyakov,
  ``A Few projects in string theory,''
  arXiv:hep-th/9304146.
  %%CITATION = HEP-TH 9304146;%%
}

\lref\windingproduction{
%\BerkoozRE
  M.~Berkooz, B.~Pioline and M.~Rozali,
  ``Closed strings in Misner space,''
  JCAP {\bf 0408}, 004 (2004)
  [arXiv:hep-th/0405126];
  %%CITATION = HEP-TH 0405126;%%
  %\SheQQ
  J.~H.~She,
  ``Winding String Condensation and Noncommutative Deformation of Spacelike
  Singularity,''
  arXiv:hep-th/0512299.
  %%CITATION = HEP-TH 0512299;%%
}

%\KarczmarekPH
\lref\KarczmarekPH{
  J.~L.~Karczmarek and A.~Strominger,
  ``Closed string tachyon condensation at c = 1,''
  JHEP {\bf 0405}, 062 (2004)
  [arXiv:hep-th/0403169].
  %%CITATION = HEP-TH 0403169;%%
}

\lref\otherD{
%\MooreWP
  G.~Moore and A.~Parnachev,
  ``Profiling the brane drain in a nonsupersymmetric orbifold,''
  arXiv:hep-th/0507190;
  %%CITATION = HEP-TH 0507190;%%
  %\MelnikovHQ
  I.~V.~Melnikov and M.~R.~Plesser,
  ``The Coulomb branch in gauged linear sigma models,''
  JHEP {\bf 0506}, 013 (2005)
  [arXiv:hep-th/0501238];
  %%CITATION = HEP-TH 0501238;%%
  %\MinwallaHJ
  S.~Minwalla and T.~Takayanagi,
  ``Evolution of D-branes under closed string tachyon condensation,''
  JHEP {\bf 0309}, 011 (2003)
  [arXiv:hep-th/0307248].
  %%CITATION = HEP-TH 0307248;%%
}

\lref\NLST{%\AharonyPA
  O.~Aharony, M.~Berkooz and E.~Silverstein,
  ``Multiple-trace operators and non-local string theories,''
  JHEP {\bf 0108}, 006 (2001)
  [arXiv:hep-th/0105309];
  %%CITATION = HEP-TH 0105309;%%
%\BerkoozUG
  M.~Berkooz, A.~Sever and A.~Shomer,
  ``Double-trace deformations, boundary conditions and spacetime
  singularities,''
  JHEP {\bf 0205}, 034 (2002)
  [arXiv:hep-th/0112264];
  %%CITATION = HEP-TH 0112264;%%
  %\WittenUA
  E.~Witten,
  ``Multi-trace operators, boundary conditions, and AdS/CFT correspondence,''
  arXiv:hep-th/0112258;
  %%CITATION = HEP-TH 0112258;%%
  A.~Sever and A.~Shomer,
  ``A note on multi-trace deformations and AdS/CFT,''
  JHEP {\bf 0207}, 027 (2002)
  [arXiv:hep-th/0203168].
  %%CITATION = HEP-TH 0203168;%%
}

%\AharonyCX
\lref\AharonyCX{
  O.~Aharony, M.~Fabinger, G.~T.~Horowitz and E.~Silverstein,
  ``Clean time-dependent string backgrounds from bubble baths,''
  JHEP {\bf 0207}, 007 (2002)
  [arXiv:hep-th/0204158].
  %%CITATION = HEP-TH 0204158;%%
}

%\AharonyCX
\lref\plasmaball{
  O.~Aharony, S.~Minwalla and T.~Wiseman,
  ``Plasma-balls in large N gauge theories and localized black holes,''
  arXiv:hep-th/0507219.
  %%CITATION = HEP-TH 0507219;%%
}

\lref\offshellstrings{
%\PolchinskiJQ
  J.~Polchinski,
  ``Factorization Of Bosonic String Amplitudes,''
  Nucl.\ Phys.\ B {\bf 307}, 61 (1988);
  %%CITATION = NUPHA,B307,61;%%
  %\CohenPV
  A.~G.~Cohen, G.~W.~Moore, P.~C.~Nelson and J.~Polchinski,
  ``Semi Off-Shell String Amplitudes,''
  Nucl.\ Phys.\ B {\bf 281}, 127 (1987).
  %%CITATION = NUPHA,B281,127;%%
}

%\BirrellPK
\lref\BirrellTaylor{
  N.~D.~Birrell and J.~G.~Taylor,
  ``Analysis Of Interacting Quantum Field Theory In Curved Space-Time,''
  J.\ Math.\ Phys.\  {\bf 21}, 1740 (1980).
  %%CITATION = JMAPA,21,1740;%%
}

%\HertogRZ
\lref\HertogRZ{
  T.~Hertog and G.~T.~Horowitz,
  ``Towards a big crunch dual,''
  JHEP {\bf 0407}, 073 (2004)
  [arXiv:hep-th/0406134].
  %%CITATION = HEP-TH 0406134;%%
}

\lref\SAE{QM SAE}

\lref\TE{
%\McGreevyCI
  J.~McGreevy and E.~Silverstein,
  ``The tachyon at the end of the universe,''
  JHEP {\bf 0508}, 090 (2005)
  [arXiv:hep-th/0506130].
  %%CITATION = HEP-TH 0506130;%%
}

%\FredenhagenUT
\lref\Schomerusanomaly{
  S.~Fredenhagen and V.~Schomerus,
  ``On minisuperspace models of S-branes,''
  JHEP {\bf 0312}, 003 (2003)
  [arXiv:hep-th/0308205].
  %%CITATION = HEP-TH 0308205;%%
}

\lref\AtickWitten{
  J.~J.~Atick and E.~Witten,
``The Hagedorn Transition And The Number Of Degrees Of Freedom Of String
Theory,''
  Nucl.\ Phys.\ B {\bf 310}, 291 (1988).
  %%CITATION = NUPHA,B310,291;%%
}

%\BarbonDD
\lref\BarbonDD{
  J.~L.~F.~Barbon and E.~Rabinovici,
  ``Touring the Hagedorn ridge,''
  arXiv:hep-th/0407236.
  %%CITATION = HEP-TH 0407236;%%
}

%\deAlwisPR
\lref\DeAlwisPR{
  S.~P.~de Alwis, J.~Polchinski and R.~Schimmrigk,
  ``Heterotic Strings With Tree Level Cosmological Constant,''
  Phys.\ Lett.\ B {\bf 218}, 449 (1989).
  %%CITATION = PHLTA,B218,449;%%
}

%\KachruED
\lref\KachruED{
  S.~Kachru, J.~Kumar and E.~Silverstein,
  ``Orientifolds, RG flows, and closed string tachyons,''
  Class.\ Quant.\ Grav.\  {\bf 17}, 1139 (2000)
  [arXiv:hep-th/9907038].
  %%CITATION = HEP-TH 9907038;%%
}

%\HarveyNA
\lref\HarveyNA{
  J.~A.~Harvey, D.~Kutasov and E.~J.~Martinec,
 ``On the relevance of tachyons,''
  arXiv:hep-th/0003101.
  %%CITATION = HEP-TH 0003101;%%
}

\lref\Matrixcosmo{
%\KarczmarekPV
  J.~L.~Karczmarek and A.~Strominger,
``Matrix cosmology,''
  JHEP {\bf 0404}, 055 (2004)
  [arXiv:hep-th/0309138];
  %%CITATION = HEP-TH 0309138;%%
  J.~L.~Karczmarek and A.~Strominger,
``Closed string tachyon condensation at c = 1,''
  JHEP {\bf 0405}, 062 (2004)
  [arXiv:hep-th/0403169].
  %%CITATION = HEP-TH 0403169;%%
  J.~L.~Karczmarek, A.~Maloney and A.~Strominger,
``Hartle-Hawking vacuum for c = 1 tachyon condensation,''
  JHEP {\bf 0412}, 027 (2004)
  [arXiv:hep-th/0405092];
  %%CITATION = HEP-TH 0405092;%%
  S.~R.~Das and J.~L.~Karczmarek,
``Spacelike boundaries from the c = 1 matrix model,''
  Phys.\ Rev.\ D {\bf 71}, 086006 (2005)
  [arXiv:hep-th/0412093];
  %%CITATION = HEP-TH 0412093;%%
%\HiranoSG
  S.~Hirano,
``Energy quantisation in bulk bouncing tachyon,''
  arXiv:hep-th/0502199.
  %%CITATION = HEP-TH 0502199;%%
 }

 %\GutperleXF
\lref\StromGut{
  M.~Gutperle and A.~Strominger,
  ``Timelike boundary Liouville theory,''
  Phys.\ Rev.\ D {\bf 67}, 126002 (2003)
  [arXiv:hep-th/0301038].
  %%CITATION = HEP-TH 0301038;%%
}

%\StromingerFN
\lref\StromTak{
  A.~Strominger and T.~Takayanagi,
``Correlators in timelike bulk Liouville theory,''
  Adv.\ Theor.\ Math.\ Phys.\  {\bf 7}, 369 (2003)
  [arXiv:hep-th/0303221].
  %%CITATION = HEP-TH 0303221;%%
}

%\SchomerusVV
\lref\SchomerusVV{
  V.~Schomerus,
  ``Rolling tachyons from Liouville theory,''
  JHEP {\bf 0311}, 043 (2003)
  [arXiv:hep-th/0306026].
  %%CITATION = HEP-TH 0306026;%%
}

%\PolyakovTP
\lref\Poly{
  A.~M.~Polyakov,
  ``A few projects in string theory,''
  arXiv:hep-th/9304146.
  %%CITATION = HEP-TH 9304146;%%
}

%\AdamsSV
\lref\APS{ A.~Adams, J.~Polchinski and E.~Silverstein, ``Don't panic! Closed string tachyons in ALE
space-times,'' JHEP {\bf 0110}, 029 (2001) [arXiv:hep-th/0108075].
%%CITATION = HEP-TH 0108075;%%
}

%\AdamsRB
\lref\TFA{
  A.~Adams, X.~Liu, J.~McGreevy, A.~Saltman and E.~Silverstein,
  ``Things fall apart: Topology change from winding tachyons,''
  JHEP {\bf 0510}, 033 (2005)
  [arXiv:hep-th/0502021].
  %%CITATION = HEP-TH 0502021;%%
}

\lref\othertach{
  M.~Headrick, S.~Minwalla and T.~Takayanagi,
``Closed string tachyon condensation: An overview,''
  Class.\ Quant.\ Grav.\  {\bf 21}, S1539 (2004)
  [arXiv:hep-th/0405064];
  %%CITATION = HEP-TH 0405064;%%
  E.~J.~Martinec,
``Defects, decay, and dissipated states,''
  arXiv:hep-th/0210231.
  %%CITATION = HEP-TH 0210231;%%
}

\lref\garybubbles{
%\HorowitzVP
  G.~T.~Horowitz,
  ``Tachyon condensation and black strings,''
  JHEP {\bf 0508}, 091 (2005)
  [arXiv:hep-th/0506166].
  %%CITATION = HEP-TH 0506166;%%
  }

%\RossMS
\lref\RossMS{
  S.~F.~Ross,
  ``Winding tachyons in asymptotically supersymmetric black strings,''
  JHEP {\bf 0510}, 112 (2005)
  [arXiv:hep-th/0509066];
  %%CITATION = HEP-TH 0509066;%%
%\cite{Cardoso:2005gj}
  V.~Cardoso, O.~J.~C.~Dias, J.~L.~Hovdebo and R.~C.~Myers,
  ``Instability of non-supersymmetric smooth geometries,''
  arXiv:hep-th/0512277.
  %%CITATION = HEP-TH 0512277;%%
  }

%\CostaEJ
\lref\CostaEJ{
  M.~S.~Costa, C.~A.~R.~Herdeiro, J.~Penedones and N.~Sousa,
  ``Hagedorn transition and chronology protection in string theory,''
  arXiv:hep-th/0504102;
  %%CITATION = HEP-TH 0504102;%%
A.~Adams and A.~Maloney, in progress.
}

%\WittenGJ
\lref\WittenGJ{ E.~Witten, ``Instability Of The Kaluza-Klein Vacuum,'' Nucl.\ Phys.\ B {\bf 195}, 481 (1982).
%%CITATION = NUPHA,B195,481;%%
}

\lref\openconfine{
%\YiHD
P.~Yi, ``Membranes from five-branes and fundamental strings from Dp branes,'' Nucl.\ Phys.\ B {\bf 550}, 214
(1999) [arXiv:hep-th/9901159].
%%CITATION = HEP-TH 9901159;%%
%\BergmanXF
O.~Bergman, K.~Hori and P.~Yi, ``Confinement on the brane,'' Nucl.\ Phys.\ B {\bf 580}, 289 (2000)
[arXiv:hep-th/0002223].
%%CITATION = HEP-TH 0002223;%%
%\SenMD
A.~Sen, ``Supersymmetric world-volume action for non-BPS D-branes,'' JHEP {\bf 9910}, 008 (1999)
[arXiv:hep-th/9909062].
%%CITATION = HEP-TH 9909062;%%
}

\lref\sen{%\SenMG
A.~Sen, ``Non-BPS states and branes in string theory,'' arXiv:hep-th/9904207.
%%CITATION = HEP-TH 9904207;%%
}

\lref\branedecay{
N.~Lambert, H.~Liu and J.~Maldacena, ``Closed strings from decaying D-branes,''
arXiv:hep-th/0303139;
  J.~L.~Karczmarek, H.~Liu, J.~Maldacena and A.~Strominger,
 ``UV finite brane decay,''
  JHEP {\bf 0311}, 042 (2003)
  [arXiv:hep-th/0306132].
  %%CITATION = HEP-TH 0306132;%%
}

\lref\otherRG{
%\VafaRA
%C.~Vafa, ``Mirror symmetry and closed string tachyon condensation,''
%arXiv:hep-th/0111051;
%%CITATION = HEP-TH 0111051;%%
%\DavidVM
J.~R.~David, M.~Gutperle, M.~Headrick and S.~Minwalla, ``Closed string tachyon condensation on twisted
circles,'' JHEP {\bf 0202}, 041 (2002) [arXiv:hep-th/0111212];
%%CITATION = HEP-TH 0111212;%%
%\HeadrickHZ
M.~Headrick, S.~Minwalla and T.~Takayanagi, ``Closed string tachyon condensation: An overview,'' Class.\ Quant.\
Grav.\  {\bf 21}, S1539 (2004) [arXiv:hep-th/0405064];
%%CITATION = HEP-TH 0405064;%%
}

\lref\davetal{
%\MorrisonJA
D.~R.~Morrison and K.~Narayan, ``On tachyons, gauged linear sigma models, and flip transitions,''
arXiv:hep-th/0412337.
%%CITATION = HEP-TH 0412337;%%
}

\lref\chicago{
%\HarveyWM
 J.~A.~Harvey, D.~Kutasov, E.~J.~Martinec and G.~W.~Moore, ``Localized tachyons and RG flows,''
arXiv:hep-th/0111154;
%%CITATION = HEP-TH 0111154;%%
}

\lref\FQS{
%\FriedanXQ
D.~Friedan, Z.~Qiu and S.~H.~Shenker, ``Conformal Invariance, Unitarity And Two-Dimensional Critical
Exponents,'' Phys.\ Rev.\ Lett.\  {\bf 52}, 1575 (1984).
%%CITATION = PRLTA,52,1575;%%
}

\lref\KMS{
%\KastorEF
D.~A.~Kastor, E.~J.~Martinec and S.~H.~Shenker, ``Rg Flow In N=1 Discrete Series,'' Nucl.\ Phys.\ B {\bf 316},
590 (1989).
%%CITATION = NUPHA,B316,590;%%
}

\lref\earlierSStach{%\KachruED
S.~Kachru, J.~Kumar and E.~Silverstein, ``Orientifolds, RG flows, and closed string tachyons,'' Class.\ Quant.\
Grav.\  {\bf 17}, 1139 (2000) [arXiv:hep-th/9907038].
%%CITATION = HEP-TH 9907038;%%
}

%\ColemanBU
\lref\ColemanBU{ S.~R.~Coleman, ``Quantum Sine-Gordon Equation As The Massive Thirring Model,'' Phys.\ Rev.\ D
{\bf 11}, 2088 (1975).
%%CITATION = PHRVA,D11,2088;%%
}

%\GreeneYB
\lref\GreeneYB{ B.~R.~Greene, K.~Schalm and G.~Shiu, ``Dynamical topology change in M theory,'' J.\ Math.\
Phys.\  {\bf 42}, 3171 (2001) [arXiv:hep-th/0010207];
%%CITATION = HEP-TH 0010207;%%
}

%\CarreauIS
\lref\CarreauIS{
  M.~Carreau, E.~Farhi, S.~Gutmann and P.~F.~Mende,
  ``The Functional Integral For Quantum Systems With Hamiltonians Unbounded
  From Below,''
  Annals Phys.\  {\bf 204}, 186 (1990).
  %%CITATION = APNYA,204,186;%%
}

\lref\classtop{
%\AspinwallNU
P.~S.~Aspinwall, B.~R.~Greene and D.~R.~Morrison, ``Calabi-Yau moduli space, mirror manifolds and spacetime
topology  change in string theory,'' Nucl.\ Phys.\ B {\bf 416}, 414 (1994) [arXiv:hep-th/9309097];
%%CITATION = HEP-TH 9309097;%%
%\WittenYC
E.~Witten, ``Phases of N = 2 theories in two dimensions,'' Nucl.\ Phys.\ B {\bf 403}, 159 (1993)
[arXiv:hep-th/9301042];
%%CITATION = HEP-TH 9301042;%%
%\DistlerMK
J.~Distler and S.~Kachru, ``(0,2) Landau-Ginzburg theory,'' Nucl.\ Phys.\ B {\bf 413}, 213 (1994)
[arXiv:hep-th/9309110].
%%CITATION = HEP-TH 9309110;%%
%\DistlerBC
J.~Distler and S.~Kachru, ``Duality of (0,2) string vacua,'' Nucl.\ Phys.\ B {\bf 442}, 64 (1995)
[arXiv:hep-th/9501111].
%%CITATION = HEP-TH 9501111;%%
}

\lref\trapping{
%\SilversteinHF
%\KofmanYC
L.~Kofman, A.~Linde, X.~Liu, A.~Maloney, L.~McAllister and E.~Silverstein, ``Beauty is attractive: Moduli
trapping at enhanced symmetry points,'' JHEP {\bf 0405}, 030 (2004) [arXiv:hep-th/0403001].
%\MohauptPQ
%T.~Mohaupt and F.~Saueressig, ``Effective supergravity actions for conifold transitions,'' arXiv:hep-th/0410272.
%%CITATION = HEP-TH 0410272;%%
%\JarvQY
%L.~Jarv, T.~Mohaupt and F.~Saueressig, ``M-theory cosmologies from singular Calabi-Yau compactifications,'' JCAP
%{\bf 0402}, 012 (2004) [arXiv:hep-th/0310174].
%%CITATION = HEP-TH 0310174;%%
%%CITATION = HEP-TH 0403001;%%
%E.~Silverstein and D.~Tong, ``Scalar speed limits and cosmology: Acceleration from D-cceleration,'' Phys.\ Rev.\
%D {\bf 70}, 103505 (2004) [arXiv:hep-th/0310221].
%%CITATION = HEP-TH 0310221;%%
}

\lref\quantop{
%\GreeneHU
B.~R.~Greene, D.~R.~Morrison and A.~Strominger, ``Black hole condensation and the unification of string vacua,''
Nucl.\ Phys.\ B {\bf 451}, 109 (1995) [arXiv:hep-th/9504145];
%%CITATION = HEP-TH 9504145;%%
%\CandelasJS
P.~Candelas and X.~C.~de la Ossa, ``Comments On Conifolds,'' Nucl.\ Phys.\ B {\bf 342}, 246 (1990);
%%CITATION = NUPHA,B342,246;%%
%\KachruRS
S.~Kachru and E.~Silverstein, ``Chirality-changing phase transitions in 4d string vacua,'' Nucl.\ Phys.\ B {\bf
504}, 272 (1997) [arXiv:hep-th/9704185].
%%CITATION = HEP-TH 9704185;%%
%\GukovZG
S.~Gukov, J.~Sparks and D.~Tong, ``Conifold transitions and five-brane condensation in M-theory on Spin(7)
%manifolds,''
Class.\ Quant.\ Grav.\  {\bf 20}, 665 (2003) [arXiv:hep-th/0207244].
%%CITATION = HEP-TH 0207244;%%
}

\lref\MandelstamHB{
S.~Mandelstam,
``Soliton Operators For The Quantized Sine-Gordon Equation,''
Phys.\ Rev.\ D {\bf 11}, 3026 (1975).
%%CITATION = PHRVA,D11,3026;%%
}
%\MandelstamHB
%\ColemanBU
\lref\ColemanBU{
S.~R.~Coleman,
``Quantum Sine-Gordon Equation As The Massive Thirring Model,''
Phys.\ Rev.\ D {\bf 11}, 2088 (1975).
%%CITATION = PHRVA,D11,2088;%%
}

%\KosterlitzXP
\lref\KosterlitzXP{
J.~M.~Kosterlitz and D.~J.~Thouless,
``Ordering, Metastability And Phase Transitions In Two-Dimensional  Systems,''
J.\ Phys.\ C {\bf 6}, 1181 (1973).
%%CITATION = JPCBA,C6,1181;%%
}

%\KogutSN
\lref\KogutSN{
J.~B.~Kogut and L.~Susskind,
``Vacuum Polarization And The Absence Of Free Quarks In Four-Dimensions,''
Phys.\ Rev.\ D {\bf 9}, 3501 (1974).
%%CITATION = PHRVA,D9,3501;%%
}

%\PolyakovFU
\lref\PolyakovFU{
A.~M.~Polyakov,
``Quark Confinement And Topology Of Gauge Groups,''
Nucl.\ Phys.\ B {\bf 120}, 429 (1977).
%%CITATION = NUPHA,B120,429;%%
}

%\AhnUQ
\lref\AhnUQ{ C.~Ahn, ``Complete S Matrices Of Supersymmetric Sine-Gordon Theory And Perturbed Superconformal
Minimal Model,'' Nucl.\ Phys.\ B {\bf 354}, 57 (1991).
%%CITATION = NUPHA,B354,57;%%
}

%\ZamolodchikovXM
\lref\ZamolodchikovXM{ A.~B.~Zamolodchikov and A.~B.~Zamolodchikov, ``Factorized S-Matrices In Two Dimensions As
The Exact Solutions Of  Certain Relativistic Quantum Field Models,'' Annals Phys.\  {\bf 120}, 253 (1979).
%%CITATION = APNYA,120,253;%%
}

%\PolchinskiFN
\lref\PolchinskiFN{
J.~Polchinski,
``A Two-Dimensional Model For Quantum Gravity,''
Nucl.\ Phys.\ B {\bf 324}, 123 (1989).
%%CITATION = NUPHA,B324,123;%%
}

\lref\Xiao{
S.~Hellerman and X.~Liu,
``Dynamical dimension change in supercritical string theory,''
arXiv:hep-th/0409071.
%%CITATION = HEP-TH 0409071;%%
}

%\SaltmanJH
\lref\SaltmanJH{
A.~Saltman and E.~Silverstein,
``A new handle on de Sitter compactifications,''
arXiv:hep-th/0411271.
%%CITATION = HEP-TH 0411271;%%
}

%\ScherkTA
\lref\ScherkTA{
J.~Scherk and J.~H.~Schwarz,
``Spontaneous Breaking Of Supersymmetry Through Dimensional Reduction,''
Phys.\ Lett.\ B {\bf 82}, 60 (1979).
%%CITATION = PHLTA,B82,60;%%
}

%\HollowoodEX
\lref\HollowoodEX{
T.~J.~Hollowood and E.~Mavrikis,
``The N = 1 supersymmetric bootstrap and Lie algebras,''
Nucl.\ Phys.\ B {\bf 484}, 631 (1997)
[arXiv:hep-th/9606116].
%%CITATION = HEP-TH 9606116;%%
}

%\BajnokDK
\lref\BajnokDK{
Z.~Bajnok, C.~Dunning, L.~Palla, G.~Takacs and F.~Wagner,
``SUSY sine-Gordon theory as a perturbed conformal field theory and  finite
size effects,''
Nucl.\ Phys.\ B {\bf 679}, 521 (2004)
[arXiv:hep-th/0309120].
%%CITATION = HEP-TH 0309120;%%
}

%\FerraraJV
\lref\FerraraJV{
S.~Ferrara, L.~Girardello and S.~Sciuto,
``An Infinite Set Of Conservation Laws Of The Supersymmetric Sine-Gordon
Theory,''
Phys.\ Lett.\ B {\bf 76}, 303 (1978).
%%CITATION = PHLTA,B76,303;%%
}

%\WittenYC
\lref\WittenYC{
E.~Witten,
``Phases of N = 2 theories in two dimensions,''
Nucl.\ Phys.\ B {\bf 403}, 159 (1993)
[arXiv:hep-th/9301042].
%%CITATION = HEP-TH 9301042;%%
}

%\VafaRA
\lref\VafaRA{
C.~Vafa,
``Mirror symmetry and closed string tachyon condensation,''
arXiv:hep-th/0111051.
%%CITATION = HEP-TH 0111051;%%
}

%\MorrisonJA
\lref\MorrisonJA{
D.~R.~Morrison and K.~Narayan,
``On tachyons, gauged linear sigma models, and flip transitions,''
arXiv:hep-th/0412337.
%%CITATION = HEP-TH 0412337;%%
}

%\BuscherQJ
\lref\BuscherQJ{
T.~H.~Buscher,
``Path Integral Derivation Of Quantum Duality In Nonlinear Sigma Models,''
Phys.\ Lett.\ B {\bf 201}, 466 (1988).
%%CITATION = PHLTA,B201,466;%%
}

%\RocekPS
\lref\RocekPS{
M.~Rocek and E.~Verlinde,
``Duality, quotients, and currents,''
Nucl.\ Phys.\ B {\bf 373}, 630 (1992)
[arXiv:hep-th/9110053].
%%CITATION = HEP-TH 9110053;%%
}

%\MorrisonYH
\lref\MorrisonYH{
D.~R.~Morrison and M.~R.~Plesser,
``Towards mirror symmetry as duality for two dimensional abelian gauge
theories,''
Nucl.\ Phys.\ Proc.\ Suppl.\  {\bf 46}, 177 (1996)
[arXiv:hep-th/9508107].
%%CITATION = HEP-TH 9508107;%%
}

%\HoriKT
\lref\HoriKT{
K.~Hori and C.~Vafa,
``Mirror symmetry,''
arXiv:hep-th/0002222.
%%CITATION = HEP-TH 0002222;%%
}

%\AharonySX
\lref\AharonySX{
O.~Aharony, J.~Marsano, S.~Minwalla, K.~Papadodimas and M.~Van Raamsdonk,
``The Hagedorn/deconfinement phase transition in weakly coupled large N gauge
theories,''
arXiv:hep-th/0310285.
%%CITATION = HEP-TH 0310285;%%
}

%\ZamolodchikovGT
\lref\ZamolodchikovGT{
A.~B.~Zamolodchikov,
``'Irreversibility' Of The Flux Of The Renormalization Group In A 2-D Field
Theory,''
JETP Lett.\  {\bf 43}, 730 (1986)
[Pisma Zh.\ Eksp.\ Teor.\ Fiz.\  {\bf 43}, 565 (1986)].
%%CITATION = JTPLA,43,730;%%
}

%\BanksQS
\lref\BanksQS{
T.~Banks and E.~J.~Martinec,
``The Renormalization Group And String Field Theory,''
Nucl.\ Phys.\ B {\bf 294}, 733 (1987).
%%CITATION = NUPHA,B294,733;%%
}

%\DeAlwisKP
\lref\DeAlwisKP{
  S.~P.~De Alwis and A.~T.~Flournoy,
``Closed string tachyons and semi-classical instabilities,''
  Phys.\ Rev.\ D {\bf 66}, 026005 (2002)
  [arXiv:hep-th/0201185].
  %%CITATION = HEP-TH 0201185;%%
}

%\LiuFT
\lref\LMS{
  H.~Liu, G.~W.~Moore and N.~Seiberg,
  ``Strings in a time-dependent orbifold,''
  JHEP {\bf 0206}, 045 (2002)
  [arXiv:hep-th/0204168];
  %%CITATION = HEP-TH 0204168;%%
%\CornalbaKD
%\lref\CornalbaKD{
and see references in
  L.~Cornalba and M.~S.~Costa,
``Time-dependent orbifolds and string cosmology,''
  Fortsch.\ Phys.\  {\bf 52}, 145 (2004)
  [arXiv:hep-th/0310099].
  %%CITATION = HEP-TH 0310099;%%
}
%\HorowitzMW
\lref\GaryJoeetal{
  G.~T.~Horowitz and J.~Polchinski,
  ``Instability of spacelike and null orbifold singularities,''
  Phys.\ Rev.\ D {\bf 66}, 103512 (2002)
  [arXiv:hep-th/0206228];
  %%CITATION = HEP-TH 0206228;%%
%\LawrenceAJ
%\lref\LawrenceAJ{
  A.~Lawrence,
  ``On the instability of 3D null singularities,''
  JHEP {\bf 0211}, 019 (2002)
  [arXiv:hep-th/0205288];
  %%CITATION = HEP-TH 0205288;%%
  M.~Fabinger and J.~McGreevy,
  ``On smooth time-dependent orbifolds and null singularities,''
  JHEP {\bf 0306}, 042 (2003)
  [arXiv:hep-th/0206196];
  %%CITATION = HEP-TH 0206196;%%\eg\
  H.~Liu, G.~W.~Moore and N.~Seiberg,
  ``Strings in time-dependent orbifolds,''
  JHEP {\bf 0210}, 031 (2002)
  [arXiv:hep-th/0206182].
  %%CITATION = HEP-TH 0206182;%%
}

\lref\nappiwitt{
  C.~R.~Nappi and E.~Witten,
  ``A Closed, expanding universe in string theory,''
  Phys.\ Lett.\ B {\bf 293}, 309 (1992)
  [arXiv:hep-th/9206078] and citations thereof.
  %%CITATION = HEP-TH 9206078;%%
}

\lref\Steve{
  L.~Fidkowski, V.~Hubeny, M.~Kleban and S.~Shenker,
  ``The black hole singularity in AdS/CFT,''
  JHEP {\bf 0402}, 014 (2004)
  [arXiv:hep-th/0306170];
  %%CITATION = HEP-TH 0306170;%%
  P.~Kraus, H.~Ooguri and S.~Shenker,
  ``Inside the horizon with AdS/CFT,''
  Phys.\ Rev.\ D {\bf 67}, 124022 (2003)
  [arXiv:hep-th/0212277].
  %%CITATION = HEP-TH 0212277;%%
}

%\HoriKT
\lref\HoriKT{
  K.~Hori and C.~Vafa,
``Mirror symmetry,''
  arXiv:hep-th/0002222.
  %%CITATION = HEP-TH 0002222;%%
}

\lref\BHfinalstate{
  G.~T.~Horowitz and J.~Maldacena,
``The black hole final state,''
  JHEP {\bf 0402}, 008 (2004)
  [arXiv:hep-th/0310281];
  %%CITATION = HEP-TH 0310281;%%
  D.~Gottesman and J.~Preskill,
 ``Comment on 'The black hole final state',''
  JHEP {\bf 0403}, 026 (2004)
  [arXiv:hep-th/0311269].
  %%CITATION = HEP-TH 0311269;%%
}

%\KosterlitzXP
\lref\KT{
  J.~M.~Kosterlitz and D.~J.~Thouless,
``Ordering, Metastability And Phase Transitions In Two-Dimensional
Systems,''
  J.\ Phys.\ C {\bf 6}, 1181 (1973).
  %%CITATION = JPCBA,C6,1181;%%
}
%\PolyakovRR
\lref\PolyakovRR{
  A.~M.~Polyakov,
``Interaction Of Goldstone Particles In Two-Dimensions. Applications To
Ferromagnets And Massive Yang-Mills Fields,''
  Phys.\ Lett.\ B {\bf 59}, 79 (1975).
  %%CITATION = PHLTA,B59,79;%%
}

%\NovikovAC
\lref\NovikovAC{
  V.~A.~Novikov, M.~A.~Shifman, A.~I.~Vainshtein and V.~I.~Zakharov,
 ``Two-Dimensional Sigma Models: Modeling Nonperturbative Effects Of Quantum
 Chromodynamics,''
  Phys.\ Rept.\  {\bf 116}, 103 (1984).
%  [Sov.\ J.\ Part.\ Nucl.\  {\bf 17}, 204.1986\ FECAA,17,472 (1986\ FECAA,17,472-545.1986)].
  %%CITATION = PRPLC,116,103;%%
}

%\FateevBV
\lref\FateevBV{
  V.~A.~Fateev and A.~B.~Zamolodchikov,
  ``Integrable perturbations of Z(N) parafermion models and O(3) sigma model,''
  Phys.\ Lett.\ B {\bf 271}, 91 (1991).
  %%CITATION = PHLTA,B271,91;%%
}

%\KlusonXN
\lref\KlusonXN{
  J.~Kluson,
  ``The Schrodinger wave functional and closed string rolling tachyon,''
  Int.\ J.\ Mod.\ Phys.\ A {\bf 19}, 751 (2004)
  [arXiv:hep-th/0308023].
  %%CITATION = HEP-TH 0308023;%%
}

%\TseytlinMZ
\lref\TseytlinMZ{
  A.~A.~Tseytlin,
``On The Structure Of The Renormalization Group Beta Functions In A Class Of
Two-Dimensional Models,''
  Phys.\ Lett.\ B {\bf 241}, 233 (1990).
  %%CITATION = PHLTA,B241,233;%%
}

%\SchmidhuberBV
\lref\SchmidhuberBV{
  C.~Schmidhuber and A.~A.~Tseytlin,
``On string cosmology and the RG flow in 2-d field theory,''
  Nucl.\ Phys.\ B {\bf 426}, 187 (1994)
  [arXiv:hep-th/9404180].
  %%CITATION = HEP-TH 9404180;%%
}

%\MosheXN
\lref\MosheXN{
  M.~Moshe and J.~Zinn-Justin,
``Quantum field theory in the large N limit: A review,''
  Phys.\ Rept.\  {\bf 385}, 69 (2003)
  [arXiv:hep-th/0306133].
  %%CITATION = HEP-TH 0306133;%%
}

%\BerkoozRE
\lref\BerkoozRE{
  M.~Berkooz, B.~Pioline and M.~Rozali,
``Closed strings in Misner space: Cosmological production of winding
strings,''
  JCAP {\bf 0408}, 004 (2004)
  [arXiv:hep-th/0405126].
  %%CITATION = HEP-TH 0405126;%%
}

%\CohenSM
\lref\CohenSM{
  A.~G.~Cohen, G.~W.~Moore, P.~Nelson and J.~Polchinski,
``An Off-Shell Propagator For String Theory,''
  Nucl.\ Phys.\ B {\bf 267}, 143 (1986).
  %%CITATION = NUPHA,B267,143;%%
}

%\SeibergHR
\lref\SeibergHR{
  N.~Seiberg,
``From big crunch to big bang - is it possible?,''
  arXiv:hep-th/0201039.
  %%CITATION = HEP-TH 0201039;%%
}

%\ZamolodchikovAH
\lref\ZZ{
  A.~B.~Zamolodchikov and A.~B.~Zamolodchikov,
``Liouville field theory on a pseudosphere,''
  arXiv:hep-th/0101152.
  %%CITATION = HEP-TH 0101152;%%
}

 %\Hikida
\lref\Hikida{
  Y.~Hikida and T.~Takayanagi,
``On solvable time-dependent model and rolling closed string tachyon,''
 Phys.\ Rev.\ D {\bf 70}, 126013 (2004)
  [arXiv:hep-th/0408124].
  %%CITATION = HEP-TH 0408124;%%
}

%\AdamsSV
\lref\AdamsSV{
  A.~Adams, J.~Polchinski and E.~Silverstein,
  ``Don't panic! Closed string tachyons in ALE space-times,''
  JHEP {\bf 0110}, 029 (2001)
  [arXiv:hep-th/0108075].
  %%CITATION = HEP-TH 0108075;%%
}
%\KrausHV
\lref\KrausHV{
  P.~Kraus, F.~Larsen and S.~P.~Trivedi,
  ``The Coulomb branch of gauge theory from rotating branes,''
  JHEP {\bf 9903}, 003 (1999)
  [arXiv:hep-th/9811120].
  %%CITATION = HEP-TH 9811120;%%
}

%\DanielssonFA
\lref\DanielssonFA{
  U.~H.~Danielsson, E.~Keski-Vakkuri and M.~Kruczenski,
  ``Black hole formation in AdS and thermalization on the boundary,''
  JHEP {\bf 0002}, 039 (2000)
  [arXiv:hep-th/9912209].
  %%CITATION = HEP-TH 9912209;%%
}

%\GiddingsZU
\lref\GiddingsZU{
  S.~B.~Giddings and S.~F.~Ross,
  ``D3-brane shells to black branes on the Coulomb branch,''
  Phys.\ Rev.\ D {\bf 61}, 024036 (2000)
  [arXiv:hep-th/9907204].
  %%CITATION = HEP-TH 9907204;%%
  }

%\HorowitzHA
\lref\HorowitzHA{
  G.~T.~Horowitz and R.~C.~Myers,
  ``The AdS/CFT correspondence and a new positive energy conjecture for
  general relativity,''
  Phys.\ Rev.\ D {\bf 59}, 026005 (1999)
  [arXiv:hep-th/9808079].
  %%CITATION = HEP-TH 9808079;%%
  }

 %\WittenZW
\lref\WittenZW{
  E.~Witten,
  ``Anti-de Sitter space, thermal phase transition, and confinement in  gauge
  theories,''
  Adv.\ Theor.\ Math.\ Phys.\  {\bf 2}, 505 (1998)
  [arXiv:hep-th/9803131].
  %%CITATION = HEP-TH 9803131;%%
  }

  %\ConstableGB
\lref\ConstableGB{
C.~Csaki, H.~Ooguri, Y.~Oz and J.~Terning,
  ``Glueball mass spectrum from supergravity,''
  JHEP {\bf 9901}, 017 (1999)
  [arXiv:hep-th/9806021];
  %%CITATION = HEP-TH 9806021;%%
  N.~R.~Constable and R.~C.~Myers,
  ``Spin-two glueballs, positive energy theorems and the AdS/CFT
  correspondence,''
  JHEP {\bf 9910}, 037 (1999)
  [arXiv:hep-th/9908175].
  %%CITATION = HEP-TH 9908175;%%
}

\Title{\vbox{\baselineskip12pt\hbox{} \hbox{SU-ITP-06/01}\hbox{SLAC-PUB-11616}}} {\vbox{ \centerline{The Inside
Story: } \centerline{ Quasilocal Tachyons
and Black Holes
%Time Dependent Field Theory and the Tachyon Final State
%
%The Infinite Probability Drive
%Mostly Harmless
%
%The Beginning of the End
%A Happy Ending
%...Finale
%
}}}
\bigskip
\centerline{Gary T. Horowitz$^1$ and Eva Silverstein$^2$}
\bigskip
\centerline{\it $^1$Department of Physics, UCSB, Santa Barbara, CA 93106}
%\smallskip
\centerline{{\it $^2$SLAC and Department of Physics, Stanford University, Stanford, CA 94305-4060}}
\bigskip
\bigskip
\noindent

We analyze the fate of excitations in regions of closed string tachyon condensate, a question crucial for
understanding unitarity in a class of black holes in string theory. First we introduce a simple new example of
{\it quasilocal} tachyon condensation in a globally stable AdS/CFT background, and review tachyons' appearance
in black hole physics.  Then we calculate forces on particles and fields in a tachyon phase using a field
theoretic model with spatially localized exponentially growing time dependent masses.  This model reveals two
features, both supporting unitary evolution in the bulk of spacetime. First, the growing energy of fields
sourced by sets of (real and virtual) particles in the tachyon phase yields outward forces on them, leaving
behind only combinations which do not source any fields. Secondly, requiring the consistency of perturbative
string theory imposes cancellation of a BRST anomaly, which also yields a restricted set of
states. Each of these effects supports the notion of a black hole final state arising from string-theoretic
dynamics replacing the black hole singularity.

\bigskip
\Date{January 2006}
%\draftmode

\newsec{Introduction}

There has been recent progress in understanding closed string tachyon condensation and applying it to problems
of gravitational interest (\eg\ \refs{\TFA,\garybubbles,\RossMS,\TE,\otherT,\cosmoT}).  New applications to
cosmological and black hole singularities motivate further analysis of the tachyonic phase in order to address
basic questions such as unitarity and the process by which black holes explode at the end of Hawking
evaporation.

As a time dependent system, a string background containing closed string tachyon condensation has no {\it a
priori} preferred vacuum state.  A string-theoretic version of a Euclidean vacuum is the simplest to control in
perturbative string theory \refs{\StromTak,\TE} via its relation to Liouville field theory.  However, in many
applications, it is important to understand if other vacua are allowed.  In this work we begin a detailed study
of this question, in the process formulating some simple new examples where tachyon condensation occurs in
globally stable string backgrounds.

Most previous discussions have focused on situations where the tachyons are either localized to a small region
of  space, or occur everywhere at once as in homogeneous cosmologies. In this paper, we examine the intermediate
case where a closed string tachyon condenses over a finite region of space. We will refer to this as
``quasilocal tachyons".   This case is of considerable interest for the following reason. Analysis of closed
string tachyon condensation yields results indicating that the process smoothly ends spacetime
\refs{\TFA,\garybubbles,\RossMS,\TE}.
%This is directly analogous to the results for  open string tachyons.
%Condensing those tachyons removes the D-branes on which open strings propagate. Similarly, condensing closed
%string tachyons seems to  remove spacetime itself - the arena on which closed strings propagate.
This suggests a perturbative string-catalyzed resolution of spacelike singularities, but raises simple questions
about unitarity in string-corrected gravity. For example, if  tachyon condensation  starts in a finite region of
space, what happens to a particle sent into this region?  Is the S matrix obtained in the bulk spacetime
unitary? Since tachyon condensation is a process which can happen in globally stable systems with asymptotic
supersymmetry, and in gravity duals to self contained field theories, these are sharp questions arising in a
wide class of backgrounds.

The basic possibilities are twofold:
\smallskip

\item{1.} Multiple states exist in the tachyon condensate phase, entangled with the outside.  Since the bulk spacetime does not include the tachyon condensate, this would yield a
failure of  spacetime unitarity.  In the context of AdS/CFT examples, this would mean that the field theory is dual
to a system consisting of more than the bulk spacetime string background.

\item{2.}  Only a single state is allowed in the tachyon condensate phase in a given system. Evolution in
the bulk spacetime remains unitary.

\smallskip
\noindent We will present evidence in favor of the second possibility.

%To understand this better, it is clearly advantageous to remove the complications associated with Hawking
%radiation and event horizons, and focus on simpler examples of quasilocal tachyon condensation.
%% I removed the following because of repetitiveness:
%These questions bear an obvious similarity to those surrounding  black hole evaporation. In fact, as discussed
%in \garybubbles\ and reviewed below, there are examples of black holes where tachyon condensation occurs inside
%the horizon before the curvature becomes large. It follows that the physics of the tachyon condensate governs
%the physics inside these black holes.  It is also important to understand the effect of quasilocal tachyon
%condensation on bulk spacetime unitarity in gravitational systems without event horizons.

We start by introducing a new example of quasilocal tachyon condensation, arising from a moving shell of D3-branes
in the gravity dual of gauge theory on a Scherk-Schwarz circle.  In this example, at low energies a
three-dimensional confining theory is induced time dependently. In the bulk, the  region inside the shell is
excised by a
tachyon condensation process and the resulting  spacetime is the AdS soliton \refs{\WittenZW,\HorowitzHA}\
describing the confined phase of the system.  This raises the question of whether information can get stuck in the
tachyon condensate phase which replaces the region inside the shell.  Similarly, in certain black holes in
string theory a tachyon phase replaces the singularity, and the information loss problem can be translated into
a question of whether the tachyon condensate phase supports a sector of states entangled with the states in the
bulk spacetime.

In order to gain some intuition for the dynamics of particles in the tachyon condensate phase, we next set up
and analyze a pure quantum field theory system with some of the same features. Tachyon condensation is described
on the worldsheet by a timelike Liouville theory with a semiclassical action deformed by the tachyon vertex
operator $\int \mu^2 {\cal O} e^{2 \kappa X^0}$, where ${\cal O}$ is an operator of dimension $\Delta <2$ and
$\kappa^2=2-\Delta$. In the analogue quantum field theory problem, the corresponding deformation of the
worldline action constitutes a space-time dependent mass squared which grows exponentially as a function of
time.  This suggests that tachyon condensation lifts closed string modes and ultimately spacetime itself, and
indeed basic amplitudes in the perturbative string theory are smoothed out by the tachyon term and produce
explicit results similar to those of the corresponding field theory analogue \refs{\StromTak,\TE}.

In ordinary quantum field theory in Minkowski space, there are two asymptotic regions (in the massless case,
past and future null infinity ${\cal J}^{\mp}$).  As we will see in detail, quantum field theory with a
localized region where the particle masses increase sufficiently rapidly with time (and homogeneously in space)
yields a new candidate asymptotic region  in which free classical particles
can get stuck and quantum mechanical wavepackets stop expanding.

This by itself raises the possibility of nontrivial states in the tachyon phase.  However, there are two
important features of the physics (also captured in the QFT model) which suggest that this is not the case.
First, would-be trapped particles generically source other fields.  These sourced fields become heavy in the
tachyon phase, leading to outward forces on such configurations.  This may lead to evacuation of any
configurations of (real or virtual) particles in the tachyon phase which source any components of the string
field.  Secondly, imposing perturbative BRST invariance forces correlations among members of the set of (real or
virtual) particles surviving in the tachyon phase (related to the analysis in \Schomerusanomaly).  Both of these
effects are reminiscent of the suggestion \BHfinalstate\ for resolving the apparent contradiction between
semiclassical calculations and unitarity of black hole evaporation; moreover the former may provide a dynamical
mechanism for satisfying the latter.

As a simple illustration, consider a particle inside the  shell of D3-branes. It is clear that the particle
cannot simply remain inside when the tachyon condenses. This is because the graviton also gets lifted in the
tachyon phase \refs{\StromTak,\TE},
 screening the energy of the particle. Since the total AdS energy must be conserved,
 the particle must either be forced out classically or quantum mechanically
 (by  pairing up with the negative energy partner of a particle creation event).

In the next section we describe the shell example in some detail, and also review the appearance of tachyons
inside black holes in both asymptotically flat and asymptotically AdS spacetimes. In section three we introduce
our quantum field theory model for the dynamics of particles and fields in the tachyon phase (which is a
generalization of \refs{\StromingerPC,\Schomerusanomaly}) and analyze in detail the behavior of excitations in
this model. We note the appearance of a BRST anomaly in the worldline description, which plausibly generalizes
to a nontrivial consistency condition in perturbative string theory. The last section contains a discussion of
applying these ideas to more realistic black holes such as Schwarzschild, and also some open questions.

\newsec{Examples of Quasilocal Tachyons}

The first example of quasilocal tachyon condensation appeared in \TFA.  There, a  winding tachyon condensation
process splits a Scherk-Schwarz cylinder in two.  This can yield decay of handles on Riemann surface target
spaces, as well as processes where the surface disconnects into multiple components. Although the discussion in
\TFA\ focused on the topology change process in the bulk of spacetime, the fact that the spatial derivatives
were required to be small means that the tachyon condensation was important over a finite region of space. Hence
this is an example of quasilocal tachyons.  However, in this example, it turns out that  small inhomogeneities in
the tachyon field alone serves to repel
most  particles from the region.

%I decided it wasn't necessary to say more about the big gradients here. I added some  %clarification after (3.11)

In this section we introduce several more examples, including ones where the tachyon gradient alone would permit
free particles to get trapped inside the tachyonic region.  In the next section we will analyze the dynamics of
 modes sent
into the tachyon phase  in this class of examples.

\subsec{Shell of D3-branes}

Our next example provides a setup where  tachyon condensation effects a time dependent transition between the
gravity dual of a compactified field theory on its Coulomb branch and  the gravity dual of a confining field
theory.  The tachyon condensate replaces the region of the spacetime which would correspond to the deep infrared
limit of the field theory.

Consider a spherical shell of D3 branes, i.e., the branes are arranged in an $S^5$ in the six dimensional  space
transverse to all the branes.  One can consider either the asymptotically flat, or asymptotically AdS versions
of this example. To use the insights from the AdS/CFT correspondence, we will focus on the asymptotically AdS
case. Configurations like this have been discussed in, e.g., \refs{\KrausHV, \GiddingsZU} (see also
\DanielssonFA\ for a different type of shell). The static metric is $AdS_5\times S^5$ outside the shell, and ten
dimensional flat space inside. Explicitly: \eqn\shell{ ds^2 = h^{-1}(r)[-dt^2  + dx_i dx^i] + h(r)[dr^2 + r^2
d\Omega_5]}
 where $i=1,2,3$ and
 \eqn\Hshell{ h(r) = {\ell^2\over r^2} \quad (r>R), \qquad h(r) = {\ell^2\over R^2} \quad (r<R)}
 As usual,  the AdS radius $\ell$ is related to the number of branes $N$ and string coupling
 $g$ via $\ell^4 = (4\pi gN)l_s^4$.
 The constant  $R$ is the coordinate radial position of the shell,
 but the above geometry is in fact independent of $R$. The proper radius of the shell is always $\ell$.
 We now compactify one of the $x_i$ with period $L$ and put antiperiodic boundary conditions for fermions.
 This identification causes the geometry to depend on $R$, as this enters into the size of the circle for
 small radius. In fact this size grows to infinity at the boundary, decreases linearly as $r$ decreases toward $R$,
 and is constant at $L R/\ell$ everywhere inside the shell  $r<R$.

 Now imagine that we add a little kinetic energy so the shell slowly contracts. To leading order in
 small velocities, the metric is given by \shell,\Hshell\ with $R$ a slowly decreasing function of time.
 Ignoring tachyons, we would eventually form a slightly nonextremal black 3-brane. But the horizon
 only arises at a very small radius depending the energy we add. Long before this, the radius of
 the circle reaches the string scale inside and on the shell.
 This occurs when
\eqn\tachyon{ {LR\over \ell} = l_s}
Just outside  the shell the winding tachyon instability should cause  the circle to pinch off \TFA. This removes the shell,
and everything inside. The result is a  ``bubble of nothing". Unlike the bubble proposed by Witten to describe
the decay of the Kaluza-Klein vacuum \WittenGJ, these bubbles can have nonzero mass and be static, rather than
expanding.
 Inside the shell there is a region of tachyon condensation along a spacelike surface (Fig. 1).  The similarity to spacetimes describing black hole evaporation is striking. This example allows us to address questions about how information gets out in a simpler setting, without the complications of Hawking radiation and large curvature.  Note that even though the low energy spacetime description is a  ``bubble of nothing", in string theory the interior  should really be thought of as the tachyon condensate.

\ifig\shellfigure{A shell of D3-branes slowly contracts. The spacetime outside is approximately $AdS_5 \times
S^5$, while the spacetime inside is approximately flat. The branes are wrapped around a Sherk-Schwarz circle,
and when this circle reaches the string scale, the winding tachyons condense. The exterior geometry becomes a
bubble  which settles down to the AdS soliton (cross $S^5$).  We will be interested in the fate of excitations
in the $<T>$ region.} {\epsfxsize3.5in\epsfbox{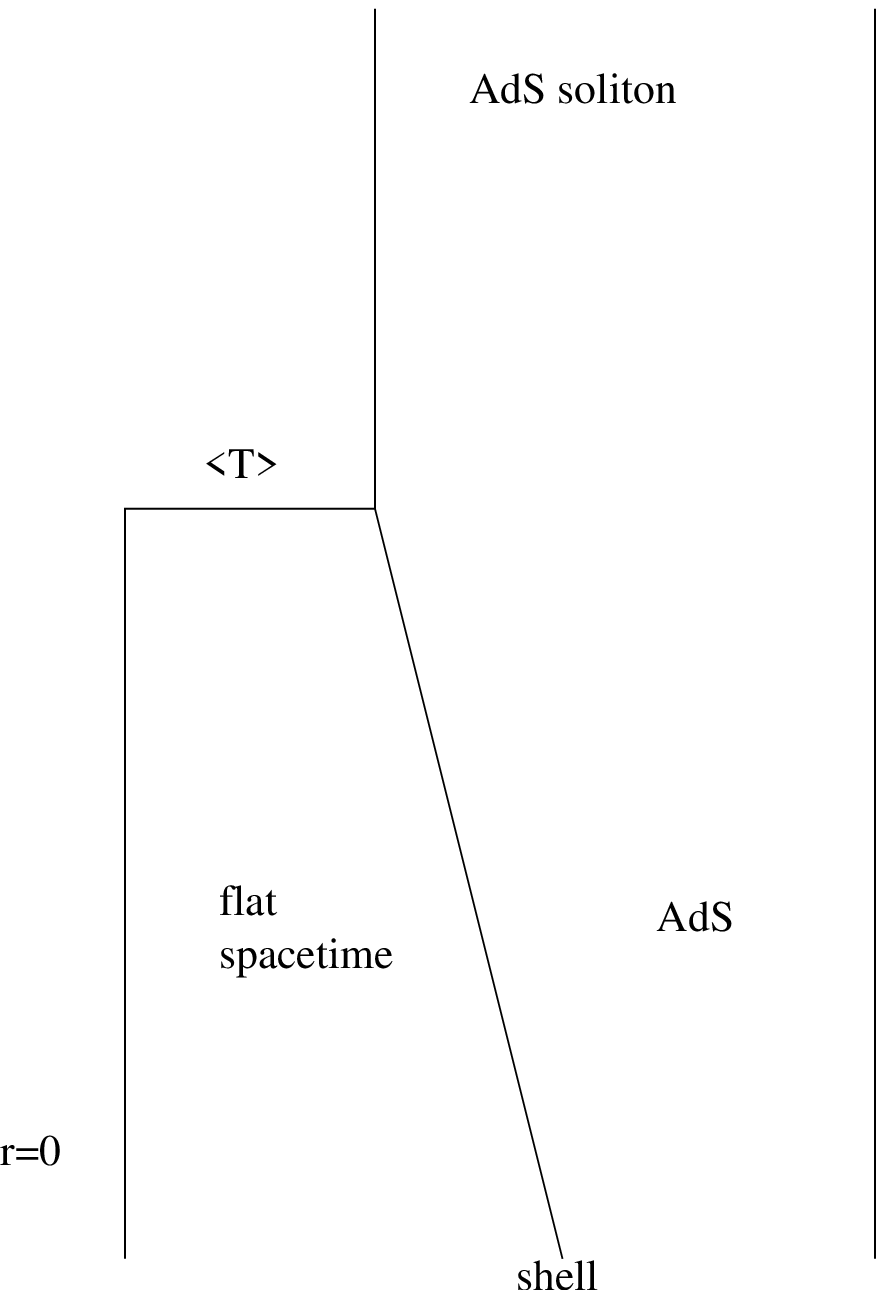}}

There is a natural candidate to describe the spacetime after the bubble settles down. This is the product of
$S^5$ and the static AdS soliton \refs{\WittenZW,\HorowitzHA}:
 \eqn\bubble{ds^2 = {r^2\over \ell^2} [-dt^2  + fd\chi^2 + dy_i dy^i] + {\ell^2 \over r^2 f} dr^2}
where
\eqn\soliton{  f(r) = 1-{r_0^4\over r^4}}
 and $ i=1,2$. The radial coordinate satisfies $r\ge r_0$ and $r=r_0$ is the bubble.
 Regularity at the bubble requires  $\chi$ to be periodic with period
 \eqn\Ldef{ L = \pi \ell^2/r_0}
  The mass of \bubble\ was computed in \HorowitzHA\ and found to be negative:
 \eqn\Esoliton{ E_{sol} = -{\pi^3 \ell^3 V_2\over 16 G_5 L^3} = -{\pi^2\over 8} {N^2 V_2\over L^3}}
 where $V_2$ is the volume of $y^i$.
 This negative energy agrees (up to a factor of $3/4$) with the Casimir energy of a weakly
 coupled ${\cal N}=4$ super Yang-Mills theory compactified on a circle of
 radius $L$ with antiperiodic fermions, though as we will discuss further below, the field theory
 is expected to confine in the infrared. Since the original spacetime had
 essentially zero energy, the transition to the bubble produces radiation in the AdS soliton background.

Most of this radiation is not produced immediately when the tachyon condenses. Instead, one first produces a
nonstatic bubble, with energy close to zero, which expands out and eventually settles down to the AdS soliton
plus radiation. To see this, let us ask what is the size of the $\chi$ circle just before it starts to pinch off
in the AdS soliton. This is given by the radius of the circle when $r$ is a few times larger than $r_0$, which
is of order $r_0L/\ell \sim \ell$. In other words, the AdS soliton \bubble\ describes a bubble in which the
circle pinches off at the AdS radius rather than the string scale.

 It is easy to write down time symmetric
initial data to describe bubbles where the circle pinches off at various radii.
Assuming the same spatial symmetries as the AdS soliton, a general four
dimensional  metric can be written in the form \eqn\newid{ ds^2 = U(r) d\chi^2 + {dr^2\over U(r) F(r)} +
{r^2\over \ell^2}dy_i dy^i } Since the extrinsic curvature is assumed to vanish, the only restriction is the
Hamiltonian constraint: \eqn\constr{ {\cal R}= -F'\left ({2U\over r}+ {U'\over 2}\right) - F\left( {2U\over r^2}
+ {4U'\over r} +  U''\right ) =  -{12\over \ell^2}  } One can pick $U(r)$ arbitrarily and solve for $F(r)$.
Since we want $U$ to vanish at $r_0$ and asymptotically be $r^2/\ell^2$, we can choose for example \eqn\Udef{
   U(r) = {r^2\over \ell^2 } - {r_0^4\over \ell^2 r^2} }
The solution for $F$ turns out to be \eqn\fsoln{
  F(r) = 1+ {b\over 3r^4 - r_0^4} }
where $b$ is an arbitrary constant. If $b=0$, this corresponds to a static slice in the AdS soliton.

Regularity at $r=r_0$ requires that $\chi$ be periodically identified with period \eqn\period{ L =\pi\ell^2
\left( r_0^2 +{b\over 2r_0^2}\right)^{-1/2}  } Using the freedom in $b$, we can obtain initial data for any $L$
and bubble location $r_0$. From \period, we simply choose \eqn\inverse{ b=  2 r_0^2\left({\pi^2\ell^4\over L^2
}-r_0^2\right) } The total energy of this initial data is easily computed with the result \eqn\Enewid{
E={V_2\over 16G_5 \ell L}\left( {r_0^4 L^2\over\pi \ell^4} - 2\pi r_0^2 \right) } Imagine fixing $L$ and varying
$r_0$. For small $r_0$ this energy is negative and decreases quadratically in $r_0$. It reaches a minimum when
$r_0 = \pi\ell^2/L$ and then increases, becoming positive and arbitrarily large at large $r_0$. The minimum
corresponds to $b=0$, consistent with the conjecture that the AdS soliton is the minimum energy solution with
these boundary conditions \HorowitzHA.

In our collapsing shell example, the circle pinches off when it reaches the string scale, which corresponds to
$r_0 = \ell l_s/L$. So its energy is only slightly negative. It will probably evolve out, oscillate around the
AdS soliton and eventually settle down.\foot{Even though the bubble will probably be momentarily at rest when it
first forms, it need not have exactly the form of $U$ chosen in \Udef. The initial data above is simply
illustrative of the general behavior.}

Inside the  shell, the geometry is similar to a spatially flat collapsing Robertson-Walker universe. However, it
is not exactly this, due to retardation effects. Let us specify that the system starts in the ground state in
the interior, and we evolve by moving the shell of branes first.  Then along a spatially flat surface, the size
of the $S^1$ will be slightly larger at the origin than near the shell. In terms of proper time $\tau$ and
proper radius $\rho$ inside the shell, the size of the circle is ${\cal L} = LR(\tau+\rho -\ell)/\ell$. So along
a constant $\tau$ surface \eqn\retardation{ {d{\cal L}\over d\rho} = {L\over \ell} {dR\over d\tau} }

We now make some comments about the dual CFT.  With supersymmetric boundary conditions, the initial shell would
be described by a point on the Coulomb branch of  $D=4$, ${\cal N} =4$ SYM. However,  since the theory is
compactified on a circle (of length $L$) with antiperiodic fermions, the fermions get masses of order $1/L$ at
tree level, and since supersymmetry is broken the scalar masses are unprotected. The Coulomb branch is lifted,
and the shell cannot stay static. However, the potential on the Coulomb branch is a small effect in our system
if we tune the bulk string coupling very small.  We are interested in a time dependent evolution (obtained by
tuning parameters and initial conditions as necessary) in which the velocity is small as the circle crosses the
string scale in proper size.  At very low energies, the theory enters a confining phase of bosonic $D=3$
Yang-Mills. This is the dual of the AdS soliton; the tachyon condensation excises the would-be IR region of the
geometry, reflecting the mass gap. Glueball masses computed from the gravity side are  of order $1/L$
\ConstableGB.

Regarding other tunable parameters in our system: instead of simply changing the shell radius, one can also
induce a tachyon transition by making the radius of the circle on which the CFT is compactified decrease with
time. In this case, the CFT is living on a time dependent spacetime (which can be chosen supersymmetric in the
far past for example). If we start in the ground state in the bulk, the information of the changing radius
propagates from the boundary to the interior, so that again the Scherk-Schwarz circle in the bulk shrinks later
near $r=0$. This again means that the tachyon turns on first near the shell and later near the origin, which
will be important in our analysis of the forces in the problem in \S3.

Similar shell examples could be constructed out of D1-D5 branes, or M2 and M5 branes. However, rather than
describe these in detail, we move on to examples involving black holes.

\subsec{Asymptotically flat black holes}

  As discussed in \garybubbles, exactly the same type of
 Scherk-Schwarz winding tachyons
appear inside some extended black holes. If a black p-brane carrying RR charge is wrapped around a circle,
the size of the circle
goes to zero at the singularity. Suppose one takes a collection of D-branes and collapses them together to form
a charged black brane. We will assume that the branes are wrapped around a circle with antiperiodic boundary
conditions. If the initial kinetic energy is very small, the situation will be similar to the shell discussed
above. The tachyon instability will set in outside the branes before the horizon is formed, and produce a
Kaluza-Klein bubble. Let us assume the initial kinetic energy is large enough to form a horizon. Then the
tachyon instability occurs inside the horizon along a spacelike surface. This can happen when the curvature is
still small so $\alpha'$ corrections are negligible, and the geometry is slowly varying so string creation
effects are small. Past this point, the evolution is dominated by the physics of tachyon condensation and no
longer given by supergravity. Outside the horizon, Hawking radiation causes the black brane to approach
extremality and the size of the circle at the horizon to shrink. When the circle reaches the string scale,
tachyon condensation will cause it to pinch off, again producing a bubble. So Hawking radiation eventually
causes all (RR charged) black branes to turn into bubbles.

As a specific example, consider the black three-brane \eqn\blackbrane{ds^2 = H^{-1/2}[-fdt^2  +  dx_i dx^i] +
H^{1/2}[f^{-1}dr^2 + r^2 d\Omega_5]} where \eqn\soliton{ H(r) = 1+ {\ell^4\over r^4}, \qquad f(r) =
1-{r_0^4\over r^4}} If we periodically identify one of the directions along the brane with period $L$ and put
antiperiodic boundary conditions, then the tachyon instability arises at a radius $r_i$  when $H(r_i)^{-1/4}L =
l_s$ which can lie inside the horizon. We now show that this can also arise when the curvature is small  and the
geometry is slowly changing. Since we want $L\gg l_s$, $H(r_i) \gg 1$, so near  $r=r_i$, $H \approx \ell^4/r^4$
and the geometry is a product of $S^5$ with radius $\ell$ and a five dimensional black brane. The circle will
reach the string scale when (cf \tachyon) $r_i =\ell l_s/L$. The Ricci curvature is of order $1/\ell^2$, but the
Weyl curvature is larger inside the horizon and of order \eqn\weyl{ C_{abcd} \sim {r_0^4\over r^4 \ell^2}} So at
the point where the tachyon instability arises, the curvature is of order $r_0^4 L^4/l_s^4 \ell^6$. For any
horizon radius $r_0$, this can clearly be made small by taking   $\ell \gg L$. The unit timelike normal to the
constant $r$ surfaces is $-f^{1/2}H^{-1/4} \partial/\partial r$. So the rate of change of the size of the circle
is \eqn\Ldot{\dot L_{S^1} = - f^{1/2}H^{-1/4} L {\partial\over \partial r} H^{-1/4} = -{(r_0^4-r^4)^{1/2} L \over
r\ell^2}} At the point where the tachyon instability begins, this gives $|\dot L_{S^1}| < r_0^2 L^2/l_s \ell^3$
which is again small whenever $\ell \gg L$.

It was shown in \garybubbles\ that to end up with a static, asymptotically flat bubble there is a restriction on
the total charge and $L$ which essentially reduces to $\ell < L$. Thus the condition to get tachyon condensation
inside the horizon with small curvature and slow time derivatives ($\ell \gg L$) is incompatible with forming a
static bubble outside the horizon. When these black branes evaporate to the point where the circle at the
horizon reaches the string scale, the resulting bubble must expand. Even when a static bubble exists as the
final endstate, there will be evolution in the bubble before it settles down, just as we saw for the shell
above.\foot{This does not contradict the condition used in \garybubbles\ that the area of the static bubble
should agree with the area of the spherical cross-section of the horizon. That corresponds to the fact that the
$S^5$ does not change during the evolution of the bubble in the shell example.}

\subsec{Asymptotically AdS black holes}

Simply dropping the one in the definition of $H$ \soliton\ converts the asymptotically flat black brane above
into an asymptotically $AdS_5\times S^5$ solution.   However
 there is an important difference between the asymptotically flat and asymptotically AdS cases. As we
 have said, in the asymptotically flat case, Hawking radiation will always cause the size of the circle
 to shrink at the horizon, so one always evolves to a bubble on the outside. In contrast, AdS acts like
 a confining box, so a typical black hole will evaporate a small fraction of its mass and quickly come
 into thermal equilibrium with its own Hawking radiation. Since tachyon condensation can still  occur
 inside, these are examples of eternal black holes with tachyon condensates inside.

Even if one lowers the temperature of the external AdS, one cannot always induce a tachyon transition outside
the horizon. For example,  the Hawking temperature of the black three-brane is $T\sim r_0/\ell^2$, and its
energy is $E_{bb} \sim N^2 T^4 V_2 L$ . The circle reaches the string scale at the horizon when $Lr_0/\ell =
l_s$. This corresponds to a temperature \eqn\tempbrane{T \sim {l_s\over \ell}{1\over L} \sim {1\over (gN)^{1/4}}
{1\over L} \ll  {1\over L}} This is a very low temperature. As discussed in \plasmaball, if we lower the
temperature of the external AdS, there is  a first order phase transition which in the CFT corresponds to a
confining/deconfining transition. In the gravity side, the transition is between the AdS soliton and the black
brane: One nucleates bubbles of the soliton on the brane. This happens  when the free energy of the black brane
is equal to the free energy of a gas in the soliton with  the same total energy.  This occurs when $E_{bb}
\sim |E_{sol}|$ which implies \Esoliton
\eqn\energyeq{ N^2 T^4 V_2 L \sim {N^2 V_2\over L^3}} That is, when $T\sim 1/L$,
which is much higher than the tachyon transition.

As another example, consider  the BTZ black hole \eqn\btz{ds^2=  - \left (r^2-r_0^2\over \ell^2\right ) dt^2 +
\left ({\ell^2 \over r^2-r_0^2}\right ) dr^2 + r^2 d\varphi^2} Since global $AdS_3$ has antiperiodic fermions
around the $\varphi$ circle, any black hole formed from collapse in $AdS_3$ must have this property. The
curvature is constant and set by $\ell$ so there are no large curvature effects provided $\ell\gg l_s$. The rate
of change of the $\varphi$ circles with respect to proper time inside the horizon  is $\dot r \approx - r_0/\ell$ so
provided $l_s\ll r_0 \ll \ell $ there are no large time derivatives and string creation effects are negligible.
Tachyon condensation will occur inside the horizon along the surface $r\sim l_s$. If the black hole can
evaporate down to $r_0\sim l_s$, then tachyon condensation at the horizon will cause the circle to pinch off and
one is left with radiation in global $AdS_3$.  However, as we have already mentioned, it is very unlikely that
the BTZ black hole will evaporate down this far.

The existence of the equilibrium configuration for black holes in AdS provides a useful simplified version of
the problem of Hawking evaporation.  If we throw a small amount of energy into the black hole, it
radiates a small amount of Hawking radiation and returns to equilibrium. Hence, this can be viewed
as an intermediate case between the shell example (with no Hawking radiation) and the asymptotically
flat black holes (with large amounts of Hawking radiation).

\subsec{AdS/CFT and Unitarity}

The examples we have discussed which are embedded in asymptotically AdS geometry benefit from the dual field
theory perspective.  This guarantees that the full system is unitary.  This alone does not guarantee that the
unitarity is maintained exclusively in the bulk spacetime; a priori it is possible that the tachyon phase
supports excitations dual to field theoretic ones.  The field theory also guarantees that energy is conserved,
and that global symmetries are respected.  These quantities are determined by graviton and gauge field behavior
at the boundary of AdS.  In this work we will focus on the direct gravity-side physics of the tachyon phase, but
will return to these constraints from AdS/CFT after some further analysis.

\newsec{Particle and Field dynamics in the $<T>$ phase}

In this section, we will gain some intuition for the dynamics of objects sent into the tachyon phase by studying
a field theory analogue.  As discussed for example in \refs{\StromTak,\TE}, the tachyon contribution to the
semiclassical worldsheet action behaves in some ways like a spacetime-dependent mass squared term for particles
in the tachyon background.  Full perturbative string computations of the partition function and of the number of
pairs of string produced in the time dependent background yield results identical to those of a field theory
with spacetime dependent masses.  This suggests that the corresponding field theory model is a good guide to the
dynamics, and we will analyze the forces on particles in the presence of fields with a spacetime dependent mass.
To start, we will discuss free particles, and then generalize to the case of more direct interest in which the
particles source fields which are themselves becoming heavy in a spacetime-dependent way.

Consider a scalar field theory on $d$-dimensional Minkowski space with a space-time dependent mass.  Its
Lagrangian density is
\eqn\lagrangian{{\cal L} = -(\del\phi)^2-m^2(\vec x, x^0)\phi^2 + {\cal L}_{interaction}}
We will be interested in the case where the mass is of the following form, and constant except in a finite
region of space
\eqn\genm{m^2(x^0, \vec x)=M^2(x^0)f(r)+m_0^2}
where $f$ has support in a finite region $r<\tilde L$, and $M\to\infty$ as $x^0\to\infty$.   A case of
particular interest will be one inspired by tachyon condensation, where
\eqn\expmass{M_T^2(x^0)=\mu^2e^{2\kappa x^0}}
but we will consider the problem in more generality. Several important features will depend on whether $M(x^0)$
grows to infinity faster or slower than linearly in $x^0$.

We will consider both the second quantized description of this system via \lagrangian\ and a first quantized
description via the worldline action
\eqn\worldlineac{{\cal S}_{wl} = \int d\tau \sqrt{g_{00}} ~ {1\over 2}\left(-g^{00}(\dot x^0)^2+g^{00}(\dot{\vec
x})^2 - m^2(x^0, \vec x) \right) }
which is a functional of the embedding coordinates $x^\mu(\tau)$ and the worldline metric $g_{00}(\tau)$.
Integrating over the worldline metric yields the equivalent form
\eqn\worldlineII{\tilde {\cal S}_{wl}= \int dx^0 ~ m(\vec x, x^0)\sqrt{1-\left({{d\vec x}\over {dx^0}}\right)^2}}
which is useful classically.  In the quantum theory, we may employ a BRST quantization to obtain an action
\eqn\BRSTac{{\cal S}_{BRST}=\int d\tau \left({1\over 2}\dot x^\mu \dot x_\mu-{1\over 2}m^2(x^0, \vec x)-\dot
b c\right)}
in terms of a standard $b-c$ ghost system.

\subsec{Particle mechanics}

Let us start by determining the trajectories of free particles, focusing on their behavior in the $r<\tilde L$
region. Starting from \worldlineII\ we find the equation of motion
\eqn\xeom{ {d\over{dx^0}}{{m {{d\vec x}\over dx^0}}\over\sqrt{1-\left({{d\vec x}\over {dx^0}}\right)^2}} + {\del
m\over{\del \vec x}}\sqrt{1-\left({{d\vec x}\over {dx^0}}\right)^2}= 0 }
To begin, let us take $f(r)$ to be constant in the $r<\tilde L$ region.   Then in this region the solutions
satisfy
\eqn\pconst{{{m {{d\vec x}\over dx^0}}\over\sqrt{1-\left({{d \vec x}\over {dx^0}}\right)^2}}=p}
for constant momentum $p$.  This yields ${{d\vec x}\over {d x^0}}=\pm p/\sqrt{m^2(x^0, \vec x)+p^2}$ from which
we learn that the distance travelled by a particle of momentum $p$ in the $r<\tilde L$ region is
\eqn\distance{|\Delta \vec x|=p \int_{t_1}^{t_2} dx^0{1\over\sqrt{m^2(x^0, \vec x)+p^2}} }
If $m$ grows faster than linearly with $x^0$ at large $x^0$, then this distance $|\Delta \vec x|$ is finite even
if we allow infinite time $t_2\to\infty$. Otherwise, the total distance diverges as $t_2\to\infty$ from the
large $x^0$ end of the integral.

In the case of exponential growth \expmass, the distance travelled within the region $r<\tilde L$ from time
$t_1=0$ to $t_2=\infty$ is
\eqn\totaldist{|\Delta \vec x| = {p\over {\omega_0\kappa}} {\rm ArcSinh}\left({\omega_0\over\mu}\right)}
where $\omega_0=\sqrt{p^2+m_0^2}$.  This grows like $(1/\kappa){\rm log}(\omega_0/\mu)$ for large $\omega_0$.
For finite $\tilde L$, there is a finite window of frequency
\eqn\limitingp{\omega_0< \omega_* = \mu ~ {\rm Sinh}(\tilde
L\kappa\omega_*/\sqrt{\omega_*^2-m_0^2})\sim_{(\omega*\gg m_0)} ~ \mu {\rm Sinh}(\tilde L\kappa)}
for which a free particle sent into the $r<\tilde L$ region at $t=0$ stays inside this region.

Near $r=\tilde L$,  $f(r)$ decreases to zero. The gradient term in \xeom\ is now important since it is amplified
by the increase in the $M(t)$ factor. This prevents particles from entering the region $r<\tilde L$ at late
time.  If $f(r)$ has no local minima in the massive phase, then even particles which enter this region at early
time are simply repelled out of the massive phase before $x^0=\infty$.  In our quasilocal tachyon problems, the
challenge to unitarity arises in the case that the spatial dependence of the tachyon terms in the worldsheet
action does not suffice to classically repel all impinging free particles.  For example in the shell discussed
in \S2.1, the retardation effects produce a tachyon gradient propelling the particles deeper into the central
region.

So far this was purely classical.  In quantum mechanics, propagation of particles is described by wave packets
which have nontrivial extent in $\vec x$.  Since wavepackets typically spread out in time, one might have
thought that the probability of finding the particle in a finite region $\tilde L$ at late time would always go to
zero.  However this is not the case. Consider sending in a Gaussian wavepacket into the $r<\tilde L$ region at
time $x^0=0$.   At large $x^0$, since the mass is increasing the particle may be described to good approximation
by nonrelativistic quantum mechanics.  A simple analysis reveals that quantum mechanical wavepackets stop
spreading at late times precisely when $m$ grows faster than linearly with $x^0$.

%Finally, let us include the possibility that our particle decays out of the central region $r<\tilde L$

%For simplicity, suppose that the theory has a $\lambda \phi^3$ interaction with $\lambda$ independent of time,
%and consider the tree level decay induced by this interaction. If $n$ is the number of particles, the decay rate
%is given by
%
%\eqn\decays{\Gamma = -{d \over{dx^0}}log n = {C\over M(x^0)}}
%
%where $C$ is the Feynman amplitude $|\lambda|^2$ times a phase space integral over final 2-particle states.
%Integrating, we find that the change in the log of the number of particles over all time is
%
%\eqn\lostparticles{-\Delta (log n)= C\int^\infty {dt\over M(t)}}
%
%which is finite precisely if $M(x^0)$ grows faster than linearly with time as $x^0\to\infty$.

We have seen that if $m(x^0)$ grows faster than linearly with $x^0$ at late times,
free particles with $\omega_0<\omega_*$ \limitingp\ can get stuck in the $r<\tilde L$ region.
Next we will consider interactions, in particular coupling our
particles to fields which are also gaining mass in the central region.

\subsec{Interactions and field energy}

Now consider coupling our particle to a field $\eta$ which is also getting massive:
\eqn\interaction{- {\cal L} = (\del\phi)^2 + m^2(\vec x, x^0)\phi^2 + (\del\eta)^2+ m_\eta^2(\vec x,
x^0)\eta^2 + \lambda\eta j}
where $j$ is a current of $\phi$ particles.  For definiteness we will consider below the case where $\eta$
couples to the energy density of $\phi$ particles.
Here
\eqn\masseta{m_\eta^2=M_0^2+f(r)M^2(x^0).}
Particles of $\phi$ source $\eta$ fields. As the particle propagates into the tachyon phase, it drags its $\eta$
field along, but this field is getting heavy as well.  This contributes to the force on the configuration and
must be taken into account in determining whether nontrivial states survive in the tachyon phase.

Let us calculate the energy contained in the $\eta$ field sourced by a $\phi$ particle.  The field classically
is given by solving
\eqn\sourceq{(\del_0^2-\vec\nabla^2 + m_\eta^2(x^0,\vec x))\eta = \lambda j(x)}
with a particle source
\eqn\sourceform{j(x)\sim m(\vec x, x^0) \delta(\vec x-\vec v x^0) .}
Let us consider a massive slowly moving particle, and neglect the velocity $\vec v$ here. The energy contained
in the field $\eta$ is given by
\eqn\energyeta{E = \int d\vec x\biggl( \dot\eta^2+(\vec\nabla\eta)^2+m_\eta^2\eta^2 \biggr)}
This field energy will depend on how far inside the tachyon phase the particle source sits, and hence will lead
to additional forces beyond those obtained in \S3.1\ just from the particle mass itself.

Before analyzing this problem explicitly, let us indicate the main point.  In the far past, the particle source
has a constant mass $m_0$, and generates a field $\eta$ scaling like $e^{-M_0r}/r^{d-3}$ as a function of the
distance $r$ from the source.  As the time dependence in the masses turn on, the $\eta$ field set in place by
the source particle finds itself on the side of a rapidly growing potential hill, and oscillates rapidly about
its minimum at zero. To get an idea for its behavior, consider for simplicity a long wavelength mode of $\eta$
which is away from its minimum at $\eta=0$ when the mass begins to rapidly increase.  It solves the equation of
motion
\eqn\eometa{\ddot\eta = -m_\eta^2\eta}
At large $x^0$ this has solutions
\eqn\largeeta{\eta\sim {1\over\sqrt{2 M(x^0)}}e^{\pm i\int^{x^0}M(t')dt'}}
yielding an energy \energyeta\ scaling like $M(x^0)$ at large $x^0$.

In our case of fields sourced by a particle in the massive phase, this effect will yield a contribution to the
energy which grows rapidly in time and also increases with increasing distance of the particle in the tachyon
phase.  Forces resulting from this work to evacuate the tachyon phase of excitations
sourcing fields.

Now let us analyze this more quantitatively. We can solve \sourceq\ in terms of the retarded two point Greens
function $G_R(x,y)$. Let us work inside the tachyon phase, where the masses depend only on $x^0$.  Then
\eqn\greens{\eta(x)=i\lambda \int d^d y G_R(\vec x-\vec y, x^0, y^0)j(y)=i\lambda\int
dy^0\sqrt{m_0^2+\mu^2e^{2\kappa y^0}}G_R(\vec x,x^0,y^0)}
We can write the Greens function $G_R$ in terms of a complete basis of mode solutions $\psi_{\omega_k}^\pm$
satisfying
\eqn\evalue{(\del_0^2+M^2(x^0))\psi_{\omega_k}^\pm(x^0)= -\omega_k^2 \psi_{\omega_k}^\pm(x^0)}
and
\eqn\normftns{ \psi_{\omega_k}^\pm(x^0) \del_0 \psi_{\omega_k}^\mp(x^0)-\psi_{\omega_k}^\mp(x^0) \del_0
\psi_{\omega_k}^\pm(x^0) = \mp i}
where $\omega_k=\sqrt{k^2+M_0^2}$ and these modes reduce to the standard Fourier modes in the pretachyonic era:
\eqn\pastform{\psi_{\omega_k}^\pm(x^0) \to_{x^0\to -\infty} {1\over\sqrt{2\omega_k}}e^{\pm i\omega_k x^0} }
The retarded Greens function is
\eqn\retprop{G_R(x,y)=\theta(x^0-y^0)\int {{d^{d-1}\vec k}\over (2\pi)^{d-1}}e^{i\vec k\cdot(\vec x-\vec
y)}\biggl( \psi_{\omega_k}^+(x^0)\psi_{\omega_k}^-(y^0) - \psi_{\omega_k}^-(x^0)\psi_{\omega_k}^+(y^0) \biggr)}

The factor $\sqrt{m_0^2+\mu^2e^{2\kappa y^0}}$ appearing in \greens\ can be separated into two parts;
$\sqrt{m_0^2+\mu^2e^{2\kappa y^0}}=\mu e^{\kappa y^0}+\bigl(\sqrt{m_0^2+\mu^2e^{2\kappa y^0}}-\mu e^{\kappa
y^0}\bigr)$.  The first term here has no contribution in the far past, while the second term has its main
contribution from the $y^0<0$ regime where the source particle of mass $m_0$ sits generating its background
$\eta$ field.  Let us consider the second piece, since we are interested in the energy carried by the field
generated by the source particle as the field enters the phase of rapidly growing mass.  This gives an
approximate expression for the $\eta$ field
\eqn\fieldI{\eta(x)\sim i \lambda\int{{d^{d-1}\vec k}\over(2\pi)^{d-1}}e^{i\vec k\cdot\vec x} \biggl(
\bigl[\int_{-\infty}^0 dy^0 m_0\psi_{\omega_k}^-(y^0)\bigr]\psi_{\omega_k}^+(x^0)-\bigl[\int_{-\infty}^0 dy^0
m_0\psi_{\omega_k}^+(y^0)\bigr]\psi_{\omega_k}^-(x^0) \biggr) }
In the pre-tachyonic phase $y^0<0$, the mode solutions behave approximately as Fourier modes \pastform. Plugging
this in and doing the $y^0$ integral reduces our estimate for the field to
\eqn\almostdone{\eta(x)\sim \lambda m_0 \int{{d^{d-1}\vec k}\over(2\pi)^{d-1}}{e^{i\vec k\cdot\vec
x}\over{\omega_k^{3/2}\sqrt{2}}} \biggl(\psi_{\omega_k}^+(x^0)+\psi_{\omega_k}^-(x^0)\biggr)}

Now at large $x^0$, the wavefunctions $\psi_{\omega_k}^\pm$ scale as in WKB like
\eqn\WKBform{\psi_{\omega_k}^\pm(x^0)\to_{x^0\to\infty} {1\over\sqrt{2 M(x^0)}}e^{\pm i\int^{x^0}M(t')dt'}}
Hence the energy in the field \energyeta\ scales like
\eqn\Escaling{E\sim m_0^2 \lambda^2 M(x^0) \cos^2 \Big (\int^{x^0}M(t')dt' \Big )
\int d^{d-1}\vec x f(|\vec
x|)\biggl(\int{{d^{d-1}\vec k}\over(2\pi)^{d-1}}{e^{i\vec k\cdot\vec x}\over{\omega_k^{3/2}}}\biggr)^2  }
where in the last step we included the fact that the tachyon phase is quasilocalized, extending over a finite
range parameterized by a function $f(|\vec x|)$ of finite support \genm.

As anticipated, this energy grows rapidly with $x^0$ and increases with increasing extent of the source particle
within the tachyon phase.  This spatial dependence is particularly strong for low dimensional examples.
%such as the effectively $1+1$-dimensional cigar geometry sourced by the matter inside %a black hole.
% I TOOK THIS OUT SINCE ITS NOT CLEAR THE MASS DIVERGES IN %SCHWARZSCHILD
Again,
this effect provides forces outward for combinations of particles which source fields.
 In the context of string
theory, a similar effect will occur for each string mode sourced by particles putatively surviving in the
tachyon phase.  This may provide a powerful effect pointing toward a full evacuation of this region, though we
have not analyzed the full evolution of general particle-field configurations here.

More specifically, an energetically costly configuration of sources and fields can lower its energy by several
processes. First, there is an outward force on the configuration of particle and field due to the field energy.
The field can redistribute itself to minimize its support in the tachyon phase.  It is tempting to make an
analogy to flux tubes and expulsion of fields from confining dynamics, but so far our discussion has been purely
perturbative. Secondly, quantum mechanically the configuration can also lower its energy by producing pairs of
virtual particles which source $\eta$ with opposite signs.  One member of the pair stays inside the tachyon
phase $r<\tilde L$, pairing with the original particle to form a configuration that does not source $\eta$, and
the other member of the pair exits the region. More generally, sets of multiple particles which altogether
source no fields might survive in the central region. In the case of a field like $\eta$ which couples to
energy, this requires, as just discussed, constituent particles which are virtual (carry negative frequency). In
the next subsection we will see that indeed perturbative virtual particles do not decouple at late times in the
tachyon phase.

Before turning to that, let us make some remarks on the relation between the field energetics and the AdS/CFT
description of the system. As discussed in \S2.4, the AdS/CFT correspondence, when applicable, implies that
energy and global symmetry charges are conserved. These are measured at the boundary by the behavior of the
graviton and bulk gauge fields. The massing up of the gravitational and gauge fields in the tachyon phase can
lead to screening of these charges. So from the AdS/CFT point of view, the energy (and conserved global charges)
must end up in the bulk rather than remaining trapped in the tachyon phase. This has an important consequence:
If the CFT has a nondegenerate ground state, then there can be no zero energy excitations in the tachyon phase.
It must approach a unique state.

In the example discussed in \S2.1, the dual field theory description of the quasilocal
tachyon condensation is a time-dependent transition to a confining dual gauge theory.  Although confinement is
not explicitly understood in Yang-Mills theory, in such a transition one qualitatively expects the following
dynamics.  In the confined theory, the gauge-invariant composite glueballs arise at an energy and size scale
commensurate with the strong coupling scale of the field theory.  In our time dependent transition, the
excitations in the tachyon phase correspond to field theoretic modes at an energy scale below the mass gap. From
the dual field theory point of view we expect forces from flux tubes to dynamically force them to shrink toward
the size scale of the glueballs in the confining theory.  The forces we analyzed in this section, which act to
force excitations into the bulk gravitational solution dual to the confining geometry, may provide a
gravity-side manifestation of this phenomenon.  This effect is similar in some ways to the description of black
hole evaporation via hadronization in \plasmaball.

\subsec{A BRST anomaly and other subtleties with the S matrix}

So far we have studied the classical dynamics of particles and fields in a localized phase of rapidly growing
mass. Next we turn to an interesting subtlety with the S matrix in such a system, which translates into a
perturbative BRST consistency condition on states in the tachyon region in the string theoretic case of
interest.  We will start by explaining the main points and then delve into more of the details.

In the quantum interacting theory, time evolution produces sets of virtual particles which are not individually
on shell.  In ordinary Minkowski space field theory, a perturbative S matrix can be obtained by extracting on
shell perturbative particle poles from Fourier transforms of Greens functions $\int d^dx  e^{ip\cdot x}G(x,
y_n)$ in the regime of integration where $x^0\to\pm\infty$. If one considers field theory in a system which
lasts for a finite time,  the $x^0$ integration only goes over a finite range, and this quantity has no
poles. In our case of interest with a phase of rapidly growing mass as $x^0\to\infty$, we will see that the new
asymptotic region at $x^0\to\infty, r<\tilde L$ also does not afford perturbative asymptotic particle poles in
the S matrix.

A correlated phenomenon is the following.  In a worldline description, by varying the action one can easily show
 that the saddle point configuration of the path integral for a particle sitting in the tachyon phase has the
property that $x^0(\tau)$ reaches infinity at a finite worldline time $\tau=\tau_*$.  In the string theoretic
generalization,  the string worldsheet in conformal gauge similarly reaches $x^0=\infty$ in finite worldsheet
time $\tau=\tau_*$, which means that the worldsheet has a hole in it.
\ifig\worldsheet{In a tachyon condensate phase, the worldsheet of a string sitting in the tachyonic region
reaches $x^0=\infty$ in finite worldsheet time $\tau=\tau_*$.  This generically leads to anomalies unless the
resulting hole $A$ in the worldsheet is unitarily mapped to a hole $A^\prime$, continuing worldsheet evolution
in the directions indicated by the arrows.  In more general circumstances the hole $A^\prime$ may be replaced by
multiple holes.} {\epsfxsize3.5in\epsfbox{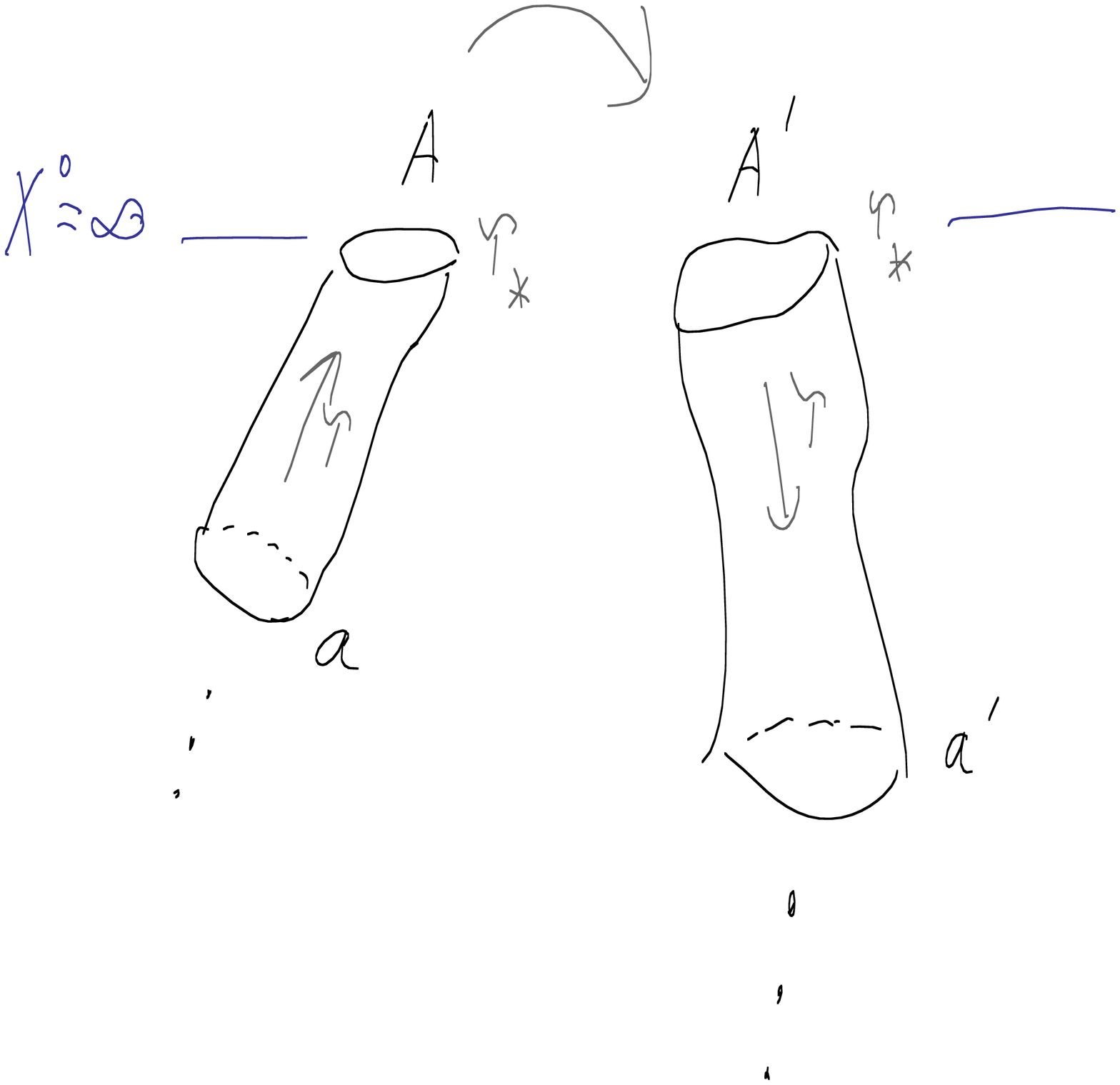}}
Worldsheets with holes are not generically BRST invariant; in special circumstances D-brane boundary states
render the holes consistent but such D-branes do not generically cancel the anomaly (as the case of the
heterotic string makes particularly clear) \offshellstrings.

One manifestation of this problem is that  the worldsheet Hamiltonian fails to be Hermitian. (We will see
explicitly below that that the worldline Hamiltonian in our field theory example is not Hermitian for general
states \Schomerusanomaly.)
% for a state consisting of a single positive
%frequency particle sitting in the phase of rapidly growing mass.  We will analyze this in detail in the field
%theory case in the rest of this subsection.  First, let us indicate briefly the main result of this for our
%question of interest.
The Hermiticity of the worldsheet Hamiltonian can be restored in the following way (see Fig. 2).  On the
worldsheet, unitary evolution persists past time $\tau_*$ if we map each hole (A) to another hole (A') by a
unitary operator on the worldsheet, and continue evolving in the direction indicated in the figure.  In real
time, this describes perturbative string (a) evolving toward the boundary correlated with another perturbative
string (a') (of negative frequency).  These two states have equal and opposite frequencies, and generically each
is individually off shell, but unitary worldsheet evolution is ensured by the correlation between the holes (A)
and (A'). Note that this correlation need not be local in space.

This is reminiscent of the black hole final state proposal
\BHfinalstate\ for solving the information puzzle. In that approach, matter and inner Hawking particles impinge on the singularity in a correlated
way.  Although it was thought that the final state would be unique, this state was not specified precisely. The proposal involves an $N\times N$
unitary matrix describing a convolution of the correlations and the bulk interactions, and any sufficiently random matrix will do.
Similarly here, imposing
cancellation of the BRST anomaly does not uniquely specify the correlations, but just requires some set of
correlations described by a unitary map.
The
simple linear relation  indicated above is too simple to model the correlations required in a real black hole, but
the same idea may be used to correlate a single worldsheet hole to multiple worldsheet holes.

We now explain in more detail the problems with Hermiticity and BRST invariance.  Consider first the worldline
description, a framework which also generalizes to the string theory case. The worldline Hamiltonian constraint
\eqn\Hconstraint{\hat H_{wl}\Psi = \left(\partial_\mu \partial^\mu
-m^2(x^0, \vec x)\right)\Psi \equiv 0}
constrains particles to lie on the mass shell.

In the case where $m$ grows faster than linearly in $x^0$, this Hamiltonian has the following property
\Schomerusanomaly. In the inner product
\eqn\naiveinner{< \psi_1 | \psi_2> \equiv \int dx^0 d^{d-1}\vec{x} \  \psi_1^*(x^0, \vec x)\psi_2(x^0, \vec x) }
$\hat H_{wl}$ is not self-adjoint on the full set of eigenfunctions of $\hat H_{wl}$.\foot{This is mathematically similar to ordinary quantum mechanics with a potential falling off faster than $-x^2$, where the Hamiltonian is again not self adjoint without extra input \CarreauIS.}  This inner product
\naiveinner\ arises both in the first quantized BRST description of the relativistic particle, and in the LSZ
prescription for the S matrix.

The failure of Hermiticity arises because of a boundary term
\eqn\boundaryterm{<\psi_1| {\partial^2\over\partial (x^0)^2} |\psi_2> - h.c. = \int d^{d-1}\vec x \left(
\psi_1^*\del_{x^0}\psi_2-  \psi_2\del_{x^0}\psi_1^*\right) |_{x^0=\infty} . }
which can be seen as follows.  The eigenfunctions $\psi_{\vec p,\Delta,\pm}$ which satisfy
\eqn\eigenvalues{\hat H_{wl}\psi_{\vec p,\Delta,\pm}=\Delta \psi_{\vec p,\Delta,\pm}}
have a WKB form valid at large $x^0$
\eqn\WKB{\psi_{p,\omega_0,\pm}|_{x^0\to\infty} \sim {1\over{\sqrt{2}}(\Delta+m^2(x^0))^{1/4}} e^{i\vec
p\cdot\vec x} e^{\pm i\int^{x^0} dt \sqrt{\Delta+m(t)^2}}}
Now the combination
\eqn\difference{<\psi_{\vec p_1,\Delta_1,\pm} \hat H_{wl} |\psi_{\vec p_2,\Delta_2,\pm}> - <\hat
H_{wl}\psi_{\vec p_1,\Delta_1,\pm} |\psi_{\vec p_2,\Delta_2,\pm}> }
would vanish if $\hat H_{wl}$ were self-adjoint.  It reduces to a boundary term of the form \boundaryterm,
evaluated at the boundaries $x^0\to\pm\infty$.  For on shell mode solutions, this boundary contribution
vanishes; on these solutions the Klein-Gordon inner product is conserved and the contributions from the future
and past boundaries cancel.

The total boundary contribution does not cancel in general for modes of different eigenvalue $\Delta$, as
explained in this context in \Schomerusanomaly.  To see this we need some regularization:  the boundary
contribution at $x^0\to -\infty$ is not well defined for general modes as the wavefunctions oscillate. Regulating
via a rescaling of t by $1-i\epsilon$ for a small real $\epsilon$, or smearing with a highly peaked distribution
in $\Delta$ as in \Schomerusanomaly, kills the contribution from the ordinary Fourier modes in the far past but
does not kill the boundary contribution in the far future. In the far future, the boundary contribution
\boundaryterm\ yields a value of $\pm 1$ for $\psi_j=\psi_{\vec p,\Delta_j,\pm}$ respectively.  Hence if we
consider all the independent eigenfunctions, the boundary term fails to cancel for this full collection of
modes.

A related point is that in the inner product \naiveinner, the full space of solutions $\psi_{\vec p,\Delta,\pm}$
do not satisfy a completeness relation:
\eqn\noncomplete{\int dx^0 \psi_{\vec p,\Delta_1, +}^*(x^0)\psi_{\vec p,\Delta_2, +}(x^0)\ne
f(\Delta_1,\Delta_2)\delta(\Delta_1-\Delta_2);}
(for any smooth function $f$); in particular this quantity does not vanish for different eigenvalues $\Delta$.
The problem can be seen from the WKB form of the wavefunctions \WKB :  for large $x^0$ these modes all approach
the same asymptotic form, leading to a failure of orthogonality in the inner product \naiveinner\ of modes of
different eigenvalues $\Delta$.  As we will see below, this behavior translates into an absence of poles in
Greens functions as the eigenvalue goes on shell $\Delta\to 0$.

In the presence of interactions, this causes problems with worldline BRST invariance. The boundary term
\boundaryterm\ violates the worldline BRST symmetry corresponding to worldline time reparameterization. This
BRST symmetry is generated by a BRST operator
\eqn\QFTBRST{Q_B = c \hat H_{wl}}
where $c$ is a Faddeev-Poppov ghost.  In the first quantized path integral, the derivation of decoupling of BRST
trivial modes depends on the Hermiticity of $\hat H_{wl}$ in the inner product \naiveinner.

The boundary term \boundaryterm\ is cancelled in states in which particles impinge on the boundary $r<\tilde L,
x^0\to\infty$ in correlated combinations with equal contributions from positive and negative frequency. Note
that the condition of cancellation of the boundary term \boundaryterm\ is a nonlocal condition; it does not
require the correlated particles at the boundary to annihilate locally in space. This is consistent with the
underlying locality in the field theory, in the same way that EPR correlations are. As discussed above, this
aspect is crucial for the application to black hole physics.

Let us discuss this issue from another point of view. In the LSZ prescription for the S matrix in Minkowski
space, asymptotic particle states are associated with poles in the Fourier transform of off shell Greens
functions with respect to momentum, for example
\eqn\LSZnormal{\int d^4 x e^{ip\cdot x}<T(\phi(x)\phi(y))>|_{p^0\to\pm\sqrt{\vec p^2+m_0^2}}\to
{\sqrt{Z}\over{(p^0)^2-\vec p^2-m_0^2}+i\epsilon} }
in an ordinary quantum field theory (without a rapidly growing mass as we have here, but instead a constant mass
$m_0$).  The poles arise from asymptotic regions $x^0\to\pm\infty$.

The analogue of this in a more general background is the convolution of the off shell Greens functions with an
eigenfunction of $\hat H\equiv -{\del^2\over{\del (x^0)^2}}+{\del^2\over{\del\vec x^2}}-M^2(x^0)$ (working
within the region $r<\tilde L$):  denote such an eigenfunction $F_\Delta(x^0)e^{i\vec p\cdot\vec x}$ where $\hat
H F=\Delta F$.  On shell modes have $\Delta=0$.  The new feature of our present case is that no such pole
appears from the $x^0\to\infty$ regime:
\eqn\LSZus{\int d^4 x F_\Delta(x)<T(\phi(x)\phi(y))>|_{\Delta\to 0} ~~ = ~~ {\rm finite} }
since in the $x^0$ integration, the region $x^0\to\infty$ is exponentially suppressed.  In this way, the system
behaves similarly to a quantum field theory living on a locally truncated Minkowski space, i.e. a space with
time stopped inside the central region.  In the latter problem as well, the worldline Hamiltonian is also not
Hermitian and no perturbative asymptotic particle states are associated with the central region.  Instead,
combinations of virtual particles impinge upon the $r<\tilde L, x^0\to\infty$ boundary in generic states.

Although this technical analysis applies most directly to the worldline quantum field theory case, similar
effects can be expected in the string theory case.  One manifestation of the problem is the holes appearing in
the worldsheet in the saddle point solutions discussed above.  Another is that unitarity relates imaginary parts
in loop diagrams arising in the regime $x^0\to\infty$ to perturbative asymptotic particle states.  The shutoff
of loop amplitudes in the Euclidean vacuum in the $x^0\to\infty$ tachyon phase suggests that no imaginary parts
will come from this regime.  Then as in field theory, perturbative asymptotic string states do not arise in the
usual way at $x^0\to\infty$ in the tachyon phase.  As discussed above, cancelling the BRST anomaly leads to
intriguing correlations at the would-be singularity reminiscent of \BHfinalstate.

%As discussed in \Schomerusanomaly, a complete basis of eigenfunctions of $\hat H$ consists of a continuous and
%discrete series of eigenfunctions, with eigenvalues $\Delta$ ranging over the set $(-\infty, 0]U_{n=0}^\infty
%\{4(\nu_0+n)^2\} $ where $\nu_0$ is an arbitrary phase.
%
%\eqn\nuvacuum{ \Psi_{\nu} = \left({\mu^2\over 4}\right)^{i\sqrt{\Delta}} \Gamma(1-i\sqrt{\Delta}) \left(
%J_{-i\sqrt{\Delta}}+{{sinh \pi(\sqrt{\Delta} -i\nu)}\over {sinh\pi(\sqrt{\Delta}+i\nu)}}J_{+i\sqrt{\Delta}}(\mu
%e^{\kappa X^0})\right) }
%
%These enter into our Greens function \greensII\ for massive fields.

\subsec{Perturbativity?}

So far we analyzed two perturbative effects following from the condensation of the perturbative string winding
tachyon. The forces coming from the field energy analyzed in \S3.2\ appear perturbatively, and work toward
evacuating any combinations of particles sourcing any components of the string field.  Worldsheet BRST
invariance is required for perturbative consistency, and is intimately connected with spacetime gauge symmetry.

In the Euclidean vacuum studied in \refs{\TE,\StromTak}, the first quantized string amplitudes are
self-consistently perturbative and calculate the components of the state in a basis of weakly coupled
multi-string states in the bulk. In more general states, it is not a priori clear if the physics remains
perturbative as combinations of (real and virtual) strings approach the singularity. In open string tachyon
problems, for example, there are indications of strong coupling physics (confinement) occurring \openconfine,
and it is tempting to speculate as many have done that an analogue happens in closed string tachyon
condensation.  This would provide its own rationale for evacuating the tachyon phase of generic excitations.
However, loop vacuum diagrams in the Euclidean vacuum provide concrete evidence for tachyon condensation
effectively massing up closed string modes \TE, which applies in particular to fluctuations of the dilaton.  If
deformations of the dilaton are indeed massed up, this might provide a mechanism for the system to remain
perturbative by freezing the dilaton at its bulk weakly coupled value. Another possible indication of
perturbativity is that D-brane probes (whose energy scales inversely with the string coupling in ordinary
spacetime string theory) are repelled from winding tachyon phases as seen in \AdamsSV\foot{See \otherD\ for
discussions of the fate of twisted D-branes in orbifolds.} (though in \KarczmarekPH\ a two-dimensional
background was studied in which nonperturbative objects were conjectured to penetrate a lightlike tachyon wall).

In any case, the basic tachyon degree of freedom driving the system away from the GR singularity is a
perturbative string mode, and as we have seen here a number of important features of the problem are accessible
perturbatively.\foot{Other approaches such as \AdSCFTsing\ may help determine the degree to which
non-perturbative corrections play a role at the singularity.}

\newsec{Discussion}

Most work on black holes in string theory, including the present work, focus on theoretical objects which are
probably not realistic. It is important to understand which techniques apply in the perhaps more physically
relevant case of Schwarzschild black holes.  In this section we assess the prospects for applying our methods in
these cases, and discuss other open problems stimulated by this work.

\subsec{Schwarzschild Black Holes}

The case of Schwarzschild black holes is of great interest.  Inside the horizon, cylinders with spherical cross
sections shrink.  Topologically stable winding tachyons thus do not appear.  However, as discussed in
\refs{\PolyakovTP,\TE}, the dynamics generating the mass gap in the two dimensional sigma model on a sphere can
behave like a superposition of winding string modes on great circles of the sphere.  A more serious challenge in
the Schwarzschild case is the rapid velocity with which the spheres shrink.

Inside a large Schwarzschild black hole in $d$ dimensions,
\eqn\BHmetric{ds^2=-\left [1-\biggl({r_0\over r}\biggr)^{d-3}\right ]dt^2+
\left [1-\biggl({r_0\over r}\biggr)^{d-3}\right ]^{-1} dr^2 +r^2d\Omega_{d-2} }
there is a $(d-2)$-sphere which starts shrinking rapidly before the spatial curvature of the sphere becomes large.
The change in sphere size $r$ with respect to proper time is
\eqn\Ldot{\dot r= - \sqrt{\biggl({r_0\over r}\biggr)^{d-3}-1} ~~~ \to ~~~ - \biggl({r_0\over r}\biggr)^{(d-3)/2} ~~
{\rm for} ~~ r\ll r_0}
which can become very rapid for $r_0\gg r\gg l_s$, i.e. while the sphere is still large.

Starting from the radius $r_c$ at which the $d$-dimensional curvature is of order $1/l_s^2$,
\eqn\curvscale{{\dot r^2\over r^2}|_{r_c}\equiv {1\over l_s^2} ~~\Rightarrow~~ r_c=\left(r_0^{d-3}l_s^2\right)^{{1\over{d-1}}}}
the time to the crunch is of order string scale.  Note that for a large black hole $r_0\gg l_s$, the spatial
curvature of the sphere is still small (the sphere is still huge).  More generally, the timescale to the crunch,
starting from a given radius $r$, is
\eqn\crunchtime{\Delta T_{crunch}={2\over d-1}r \left( {r\over r_0}\right)^{{d-3\over
2}}={2\over{d-1}}\left(r\over r_c\right)^{{{d-1}\over 2}}l_s}
This rapid velocity causes particle and string production:  in particular a simple estimate suggests that a
Hagedorn density of strings is produced by the time the sphere has shrunk to $r=r_c$.  The back reaction of this
gas of strings may ultimately behave like a winding tachyon condensate, as suggested also in \windingproduction,
but this has yet to be controlled.  More generally, the rapid shrinking of the sphere can lead to
non-adiabaticity for a large class of extended objects, whose spectrum depends on the internal degrees of
freedom of the compactification.

This situation is improved considerably in simple models of black hole evaporation. The fact that the null
energy condition is violated near the horizon (which is required in order for the area to decrease) causes the
spheres to shrink  much more slowly near the horizon. To see this, consider the Vaidya metric in four
dimensions: \eqn\Vaidya{ ds^2=-\left( 1-{2G M(v)\over r}\right) dv^2 + 2dvdr  + r^2d\Omega} This is a solution
to Einstein's equation with a null fluid source and has frequently been
 used to model an evaporating black hole. The unit timelike normal to a constant $r$ surface inside the horizon is
\eqn\normal{n= \left ({2GM(v)\over r} -1\right )^{-1/2} {\partial\over \partial v} - \left ({2GM(v)\over r }-1\right )^{1/2} {\partial\over \partial r} }
So the rate of change of the spheres is
\eqn\ratesphere{\dot r = - \left ({2GM(v)\over r} -1\right )^{1/2}}
The velocity will be less than one on a surface of constant $r$, provided $r< 2GM(v) < 2r$.
At infinity, a black hole loses mass at the rate
\eqn\massloss{ {dM\over dt} \sim - AT^4 \sim -(GM)^{-2}}
So  $G^2M^3 \sim t_0-t$. Since $v=t+$ constant along a surface at large $r$, we set
\eqn\massv{M(v) = M_0 \left (1-{v\over v_0}\right )^{1/3}}
It now follows that the proper length of the region over which the velocity is less than one
is $ r^3/G = r(r/l_p)^2$. So even when $r$ is the string scale, the distance
from the horizon over which the velocity stays small can be much greater than the string scale.

Since the velocity stays small near the horizon, the arguments of \refs{\TFA,\TE} now suggest that when the
horizon reaches the string scale, it will pinch off, removing the region of large curvature and the singularity.
This provides further support for the correspondence principle of  \corrprinc. It was argued there that
Schwarzschild black holes radiate until the curvature at the horizon reaches the string scale. At this point,
the black hole makes a transition to an excited fundamental string. Since the excited string lives in a space
which is essentially flat, the sigma model argument provides a dynamical mechanism for the transition from the
black hole to the excited string.  In order to understand the information flow however, we would need to also
control the region deeper inside the black hole where the velocity is still large in order to account for
excitations that could putatively be trapped there.

\subsec{Other Future Directions}

\subsubsec{Cosmological Case}

In this work we have focused on situations with localized tachyon condensation, or localized regions of
exponentially growing mass in the field theory model of particle and field dynamics in the tachyon phase.  In
the case of spatially delocalized tachyon condensation \refs{\TE,\cosmoT} the question of allowed states is also
of interest. Although time effectively stops in the case of decay to nothing \TE, this alone does not preclude
unitarity; ordinary quantum mechanics formulated on a finite time interval is unitary.  It is of interest to
understand whether multiple states are allowed at a cosmological singularity.  The forces we discussed which
help to evacuate the tachyon phase in the quasilocal case do not serve this role in a situation with spatially
delocalized tachyon condensation.  The perturbative BRST anomaly, when applicable, does restrict the allowed
states somewhat.  Neither of these effects is as powerful in the cosmological case as in the quasilocal case
discussed here.

\subsubsec{Worldsheet Analysis}

In our analysis we reverted to a field theory model of some of the dynamics in the tachyon phase (generalizing
that of \StromingerPC, which has withstood several tests of its applicability in the full perturbative string
theory \refs{\StromTak,\TE}). It would be advantageous to test this quasilocalized version of the field theory
model further using full worldsheet string calculations.  The prescription suggested in Fig. 2 may require
techniques such as those suggested in \refs{\offshellstrings,\NLST, \AharonyCX}\ which could provide a formalism
for describing the consistent states.

One issue for which a full worldsheet analysis is necessary is the backreaction of the large energy density
produced when the tachyon condenses. A naive supergravity analysis would indicate that this energy density
immediately produces large curvature. However, since the graviton is also becoming effectively massive,
supergravity is not a good approximation.

\subsubsec{Models and Constraints}

We end with a speculative comment on the possibility of connection to real-world astrophysics. The phenomena
discussed in \garybubbles\ show that the endpoint of Hawking evaporation can result in new types of black hole
explosions. It is of interest to translate our growing understanding of string-corrected gravity and
singularities to a theory of black hole explosions more generally.  The production of primordial black holes
small enough to evaporate in our causal past is at best a model dependent proposition, so new effects in black
hole evaporation will probably simply serve to mildly constrain model building. Still, it is interesting to
contemplate the possibility of ``fundamental" origins for astrophysically accessible bursts and jets of
energy.\foot{Amusingly, jets appear both in particle physics hadronization processes and in gamma ray bursts;
 the relation between black holes and confinement makes it tempting to seek a connection, although the
astrophysical jets are probably accounted for by effects of angular momentum.}

\bigskip
\bigskip
\centerline{\bf{Acknowledgements}}

We would like to thank John McGreevy for early collaboration and many useful discussions.  We would like to
thank D. Marolf, J. Maldacena, V. Schomerus, S. Shenker, A. Strominger, and L. Susskind for useful discussions
and O. Aharony for comments on the manuscript. We are supported in part by the DOE under contract
DE-AC03-76SF00515 and by the NSF under grants 9870115 and PHY-0244764.

\listrefs

\end